\documentclass[runningheads]{llncs}
\usepackage[T1]{fontenc}
\usepackage[table,xcdraw]{xcolor}
\usepackage{graphicx}
\usepackage{tikz}
\usepackage{etoolbox}
\usepackage{algorithmic}
\usepackage[linesnumbered,ruled,vlined]{algorithm2e}
\usepackage{adjustbox}
\usepackage{booktabs}
\usepackage{tabularx}
\usepackage{multirow}
\usepackage{makecell}
\usepackage{xurl}
\usepackage{threeparttable}

\newcommand*{\circled}[1]{%
    \tikz[baseline=(char.base)]\node[shape=circle, fill=black, text=white,
    inner sep=0pt, outer sep=0pt, minimum size=1em] (char)
    {\footnotesize #1};%
}
\robustify{\circled}

\newcommand{\consolaslike}[1]{\texttt{\textcolor[HTML]{078080}{\ttfamily#1}}}

\begin{document}
\title{Enhancing LLM-based Specification Generation via Program Slicing and Logical Deletion}
\titlerunning{Enhancing LLM-based Specification Generation}
%
%
\author{
Zehan Chen\inst{1} \and
Long Zhang\inst{3} \and
Zhiwei Zhang\inst{1} \and
JingJing Zhang\inst{4} \and
Ruoyu Zhou\inst{1} \and
Yulong Shen\inst{1} \and
JianFeng Ma\inst{2} \and
Lin Yang\inst{5}\thanks{Corresponding author.}
}
%
%
\institute{
School of Computer Science and Technology, Xidian University, Xi'an, Shannxi, China \\
\email{zehanchen@stu.xidian.edu.cn, zwzhang@xidian.edu.cn, 23031212412@stu.xidian.edu.cn, ylshen@mail.xidian.edu.cn} \and
School of Cyber Engineering, Xidian University, Xi'an, Shannxi, China \\
\email{jfma@mail.xidian.edu.cn} \and
National Key Laboratory of Science and Technology on Information System Security, AMS, Beijing, China \\
\email{zhanglong10@nudt.edu.cn} \and
Unit 32092 of PLA, Beijing, China \\
\email{lg02103@163.com} \and
National Key Laboratory of Science and Technology on Information System Security, Systems Engineering Institute, AMS, Beijing, China \\
\email{yanglin61s@126.com}
}
\maketitle              
\begin{abstract}
Traditional formal specification generation methods are typically tailored to specific specification types, and therefore suffer from limited generality. In recent years, large language model (LLM)-based specification generation approaches have emerged, offering a new direction for improving the universality of automated specification synthesis. However, when dealing with complex control flow, LLMs often struggle to precisely generate complete specifications that cover substructures. Moreover, the distinctive verification pipelines adopted by existing approaches may incorrectly discard logically correct specifications, while verification tools alone cannot reliably identify correct specifications. To address these issues, we propose SLD-Spec, an LLM-based specification generation method that combines program slicing and logical deletion. Specifically, SLD-Spec augments the conventional specification generation framework with two key stages: (1) a program slicing stage that decomposes the target function into several smaller code slices, enabling LLMs to focus on more localized semantic structures and thereby improving specification relevance and completeness; and (2) a logical deletion stage that leverages LLMs to perform logical reasoning and filtering over candidate specifications so as to retain logically correct ones. Experimental results show that SLD-Spec consistently outperforms existing methods on datasets containing programs of varying complexity, verifying more programs and generating specifications that are more relevant and more complete. Further ablation studies indicate that program slicing mainly improves specification relevance and completeness, whereas logical deletion plays a key role in increasing verification success rates.

\end{abstract}
\section{Introduction}
In program verification, formal specifications are used to precisely characterize program behaviors and the properties they are required to satisfy, thereby addressing the ambiguity that often arises when software requirements are described in natural language~\cite{Kamsties-01,Shah-01}. However, manually writing such specifications is highly labor-intensive~\cite{Li-01,Masoudi-01}. Traditional research on automated specification generation typically focuses on only specific types of specifications, which limits their generality~\cite{Ryan-01,Si-01,Yao-01}. Recent studies~\cite{Kamath-01,Ma-01,Wen-01} have begun to leverage the strong comprehension and generation capabilities of large language models (LLMs) to automatically generate a wide range of specifications for programs.

Although the introduction of LLMs addresses the issue of diversity in specification generation, this line of research has not yet been thoroughly explored and still suffers from the following three major challenges:
\begin{itemize}
    \item Complex control flow within a function makes it difficult for LLMs to accurately distinguish the scopes of different substructures, leading them to generate specifications that do not belong to the target substructure or to omit critical ones~\cite{Thanh-01,Wu-01}.
    \item Existing study~\cite{Wen-01} applies verification tools immediately after generating specifications for substructures to reduce the risk of error propagation. However, due to the lack of necessary contextual information, this practice may incorrectly eliminate specifications that are logically correct.
    \item Some studies~\cite{Kamath-01,Wen-01} leverage LLMs to generate a large number of candidate specifications to cover potential semantics and then perform filtering at later stages. However, verification tools are designed for rigorous proof checking rather than for discriminative selection. This mismatch makes the use of verification tools to filter specifications unreliable.
\end{itemize}

To address the above challenges, we introduce two core strategies: (1) Before generating specifications, we decompose functions into smaller slices. This approach reduces the amount of code provided to LLMs in each query and prevents irrelevant content from influencing specification generation. (2) After generating specifications for substructures within a function, we do not rely on verification tools to filter out specifications that fail verification. Instead, inspired by the "LLM-as-a-Judge" paradigm, we use LLMs to assess the consistency between the generated specifications and the code, thereby temporarily retaining logically correct specifications.

Based on these insights, we propose SLD-Spec, a novel LLM-based specification generation approach enhanced by program slicing and logical deletion. It extends the traditional \textit{guess-and-verify} paradigm~\cite{Kamath-01} into an automated specification generation framework consisting of four phases: \textit{slice–guess–logical delete–verify}. First, in the slicing phase, SLD-Spec performs static analysis to construct the program’s function call graph (FCG) and decomposes each function into mutually independent code slices in a bottom-up order. Second, in the guessing phase, SLD-Spec constructs prompts for each slice, incorporating a few-shot learning strategy~\cite{Brown-01}, and queries LLMs to generate a set of candidate specifications. Next, SLD-Spec replaces verification-tool-based filtering with a logical deletion phase, in which each candidate specification is evaluated and logically correct specifications are retained. Finally, in the verification phase, SLD-Spec uses verification tools to check all specifications of the entire function. If verification succeeds, all specifications are accepted as the final result; otherwise, SLD-Spec iteratively removes incorrect specifications based on the reported analysis information until the function is either verified successfully or only assertion violations remain.

To comprehensively evaluate SLD-Spec, we compare it with two state-of-the-art (SOTA) LLM-based approaches, AutoSpec~\cite{Wen-01} and SpecGen~\cite{Ma-01}, on two datasets. Specifically, on the dataset consisting of simple programs, SLD-Spec significantly increases the number of successfully verified programs, while not reducing verification efficiency despite the introduction of two processing phases. Furthermore, we construct a dataset featuring complex control flow. On this dataset, SLD-Spec substantially outperforms the other methods, generating specifications that are more numerous, more accurate, and more relevant. Finally, we conduct ablation studies on the two newly introduced phases, demonstrating the effectiveness of both program slicing and logical deletion. The contributions of this paper are as follows:
\begin{itemize}
    \item We propose SLD-Spec, a novel automated specification generation approach. It leverages program slicing to decompose complex functions into code slices, thereby improving both the quantity and relevance of specifications generated by LLMs. Moreover, logical deletion mitigates the over-pruning of logically correct specifications caused by missing contextual information. Acting as a middleware between LLMs and verification tools, SLD-Spec pre-filters candidate specifications prior to formal verification.
    \item We construct a program dataset with complex control flow. Compared with prior work, this dataset contains longer code and more intricate control structures. In addition, we extend the evaluation metrics used in previous studies. Together, the new dataset and metrics enable a more comprehensive assessment of the effectiveness and efficiency of different approaches.
    \item We compare SLD-Spec with two SOTA approaches. SLD-Spec successfully verifies 37 out of 51 programs and 10 out of 11 programs on the two datasets, respectively, outperforming the baseline approaches. In addition, ablation studies demonstrate the effectiveness of the two proposed components.
    \item All tools, datasets, and experimental results developed in this work are publicly released to support further research.
\end{itemize}

\section{Background and Motivation}
\subsection{Program Slicing}
Program slicing~\cite{Weiser-01} is a technique for automatically decomposing programs by analyzing their data and control flow. It begins with a subset of the program's behavior and simplifies the program to the minimal form, while still preserving that behavior. A slice consists of the portions of the program that may affect the values computed at points of interest, known as slicing criteria. These criteria are represented by a tuple $\langle p, V \rangle$, where $p$ denotes a specific point in the program and $V$ is a subset of the program's variables.

\subsection{Specification Language}
A specification language is a mathematically and logically grounded formalism used to formally and precisely describe a system’s requirements, behaviors, and properties. In this work, we focus on a specification language for C programs, namely the ANSI/ISO C Specification Language (ACSL)~\cite{Patrick-01}. ACSL primarily provides function contracts and loop annotations. Function contracts use the {\consolaslike{requires}}, {\consolaslike{ensures}}, and {\consolaslike{assigns}} clauses to specify preconditions, postconditions, and permissible side effects of functions. Loop annotations include {\consolaslike{loop invariant}} clauses to describe properties preserved across iterations, {\consolaslike{loop assigns}} clauses to restrict memory modifications, and {\consolaslike{loop variant}} clauses to ensure loop termination.

\subsection{Motivating Example}
\begin{figure}[tbp!]
    \centering
    \includegraphics[width=\textwidth]{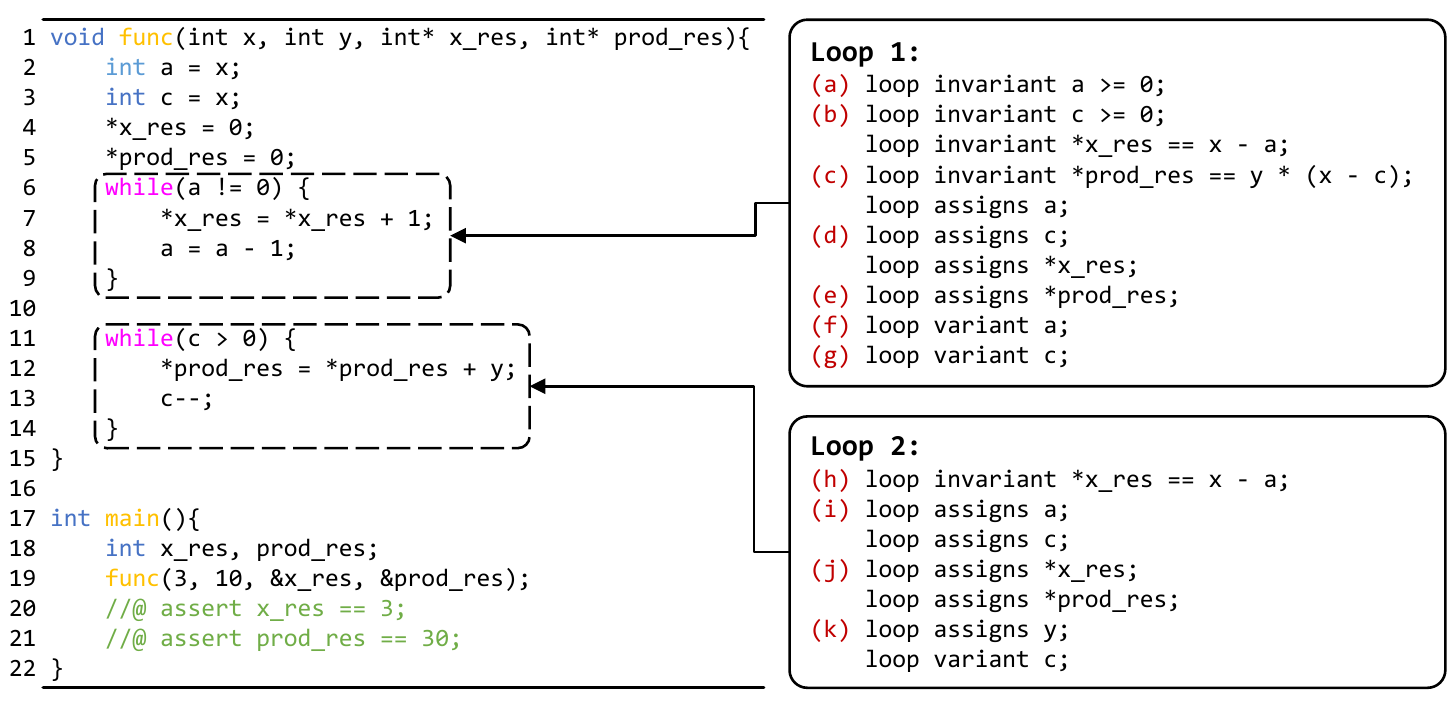}
    \caption{An Example of Specifications Generated by Existing Approaches.}
    \label{pic: motivation}
\end{figure}
Figure~\ref{pic: motivation} illustrates the specifications generated by existing approaches (right) for a function containing two loop structures (left), revealing several issues. First, when LLMs generate specifications for an individual loop, they may be influenced by irrelevant code or previously generated specifications. As a result, the generated specifications can be semantically unrelated to the corresponding loop in both Loop~1 and Loop~2, such as specifications (b)–(e) and (h)–(j). Although some of these specifications can be successfully verified, they fail to describe the actual behavior of the code and are therefore meaningless, requiring manual removal in subsequent steps. Second, the correctness of specification (a) depends on the precondition {\consolaslike{requires x >= 0}}. However, under approaches that adopt a hierarchical generation mechanism, function contracts have not yet been generated at this stage. Consequently, the verification tool deems this specification incorrect and removes it, which in turn affects the proof of other specifications in Loop~1. Finally, some studies take the union of multiple generation results as the candidate specification set. However, due to approximations used by the verification tool, a variable may still be considered valid in a {\consolaslike{loop variant}} clause even if it is not modified by the code (e.g., specification (j)). This, in turn, interferes with the proof of other specifications (e.g., the postcondition {\consolaslike{ensures *x\_res == x}}). 
Moreover, when multiple {\consolaslike{loop variant}} clauses appear in the same loop, the verification tool treats the earliest one as a syntactic error (e.g., specification (f)), rather than determining which clause is correct. Consequently, filtering specifications based on the tool’s output becomes unreliable.

Overall, these issues are not specific to particular instances but stem from inherent limitations of existing LLM-based specification generation pipelines, including context-dependent generation, hierarchical dependencies among specifications, and mismatches between the operating principles of LLMs and verification tools. These observations motivate the introduction of mechanisms to explicitly address and mitigate these deficiencies.


\section{Methodology}
\begin{figure*}[tbp!]
    \centering
    \includegraphics[width=1\textwidth]{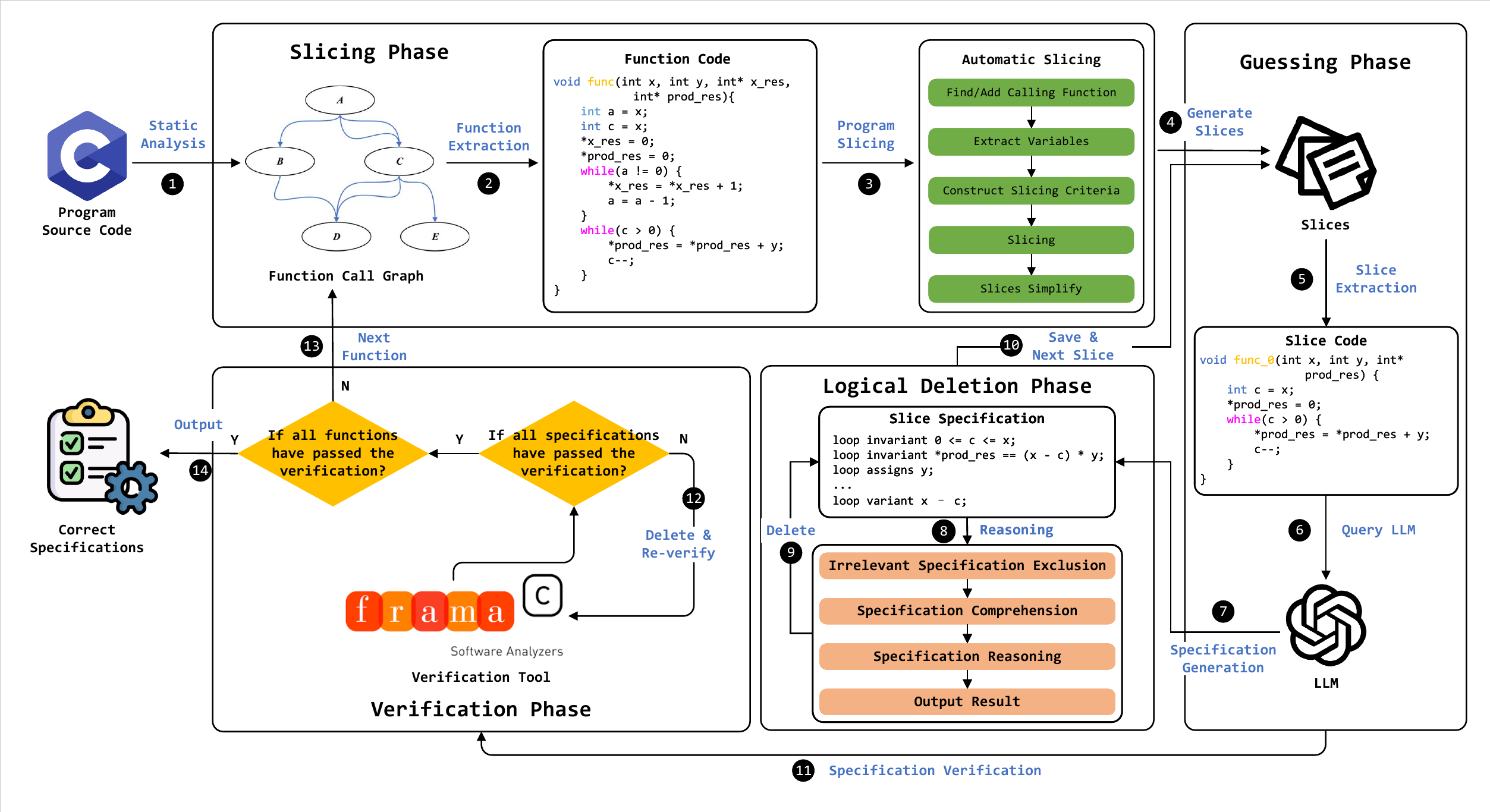}
    \caption{Overview of SLD-Spec.}
    \label{pic: Overview of SLD-Spec}
\end{figure*}

The overview of SLD-Spec is presented in Fig.~\ref{pic: Overview of SLD-Spec}, which comprises four main phases. In the slicing phase (Section~\ref{subsec: Program Slicing}), SLD-Spec first constructs the FCG of the input program through static analysis (\,\circled{1}\,). It then extracts the target function for specification generation in a bottom-up order (\,\circled{2}\,) and performs program slicing on it (\,\circled{3}\,). Since program slicing requires manual identification of variables of interest~\cite{Chalupa-01,Shahandashti-01}, we design an algorithm that automatically detects relevant variables within the function and performs slicing accordingly (\,\circled{4}\,). In the guessing phase (Section~\ref{subsec: Specification Generation}), SLD-Spec iteratively extracts an individual slice from the slice set (\,\circled{5}\,). Each slice is combined with few-shot examples to construct a prompt, which is then fed into LLMs (\,\circled{6}\,) to generate candidate specifications for the slice (\,\circled{7}\,). In the logical deletion phase (Section~\ref{subsec: Logical Deletion}), SLD-Spec conducts logical reasoning (\,\circled{8}\,) on the generated candidate specifications using a four-step process: exclusion, comprehension, reasoning, and output. Based on the results of this reasoning process, incorrect specifications are identified and removed (\,\circled{9}\,). The retained specifications are then stored, and the system proceeds to process the next slice. This continues until specifications for all slices have been generated (\,\circled{10}\,). In the verification phase (Section~\ref{subsec: Specification Verification}), SLD-Spec aggregates all retained specifications and submits them to a verification tool (\,\circled{11}\,). If any specifications fail verification, the incorrect ones are removed according to the tool's output and the verification process is repeated (\,\circled{12}\,). If verification succeeds, the system continues with specification generation for the next function (\,\circled{13}\,). This process repeats until all functions have been verified, at which point the final, verified specifications are produced as output (\,\circled{14}\,).

\subsection{Program Slicing}
\label{subsec: Program Slicing}
We adapt AutoSpec’s hierarchical generation approach~\cite{Wen-01} within SLD-Spec. Specifically, SLD-Spec constructs the abstract syntax tree (AST) by traversing the program's source code to identify the call relationships between functions, thereby creating a FCG. This process reduces the complexity of specification generation from the program level to the function level, allowing LLMs to avoid generating specifications for the entire program at once. Unlike AutoSpec, which directly generates specifications for loops and functions, we introduce the concept of slicing between them, establishing a “loop–slice–function–program” sequence for specification generation. Introducing program slicing offers two advantages: (1) it reduces the complexity of functions, thereby enabling LLMs to interpret function contents more accurately; and (2) it groups related code into the same slice, preventing LLMs from being misled by irrelevant specifications during the generation process.

Although program slicing has been extensively studied, integrating it with specification generation to achieve a fully automated workflow presents several challenges. First, some program slicing methods~\cite{Chalupa-01} require the existence of an entry function that calls the function being sliced. However, developers may only wish to verify the correctness of a function and may not provide any calling function. Second, although program slicing can accept multiple manually specified slicing criteria, it computes the set of code that is jointly affected by these criteria. In contrast, our goal is to automatically derive multiple slicing criteria from the code and partition the program into several mutually independent slices.

\begin{algorithm}[tbp!]
\scriptsize
\caption{Automatic Function Slicing}
\label{algo: Automatic Function Slicing} 
\SetKwProg{Fn}{Function}{\string:}{}
\SetKwInOut{Input}{input}
\SetKwInOut{Output}{output}
\SetKwFunction{AutoSlicing}{AutoSlicing}
\SetKwFunction{GenCallFunc}{GenCallFunc}
\SetKwFunction{GetFuncVar}{GetFuncVar}
\SetKwFunction{InsSliCrit}{InsSliCrit}
\SetKwFunction{append}{append}
\SetKwFunction{Slicing}{Slicing}
\SetKwFunction{SimplifySlicing}{SimplifySlicing}
\Input{$F$: the function to be sliced} 
\Output{$S_{fs}$: the set of functions slices}
\BlankLine 
\Fn{\AutoSlicing{F}}{
    $S_{fs} = \emptyset$ \\
    $S_{sc} = \emptyset$ \textcolor[HTML]{70ad48}{ \tcp*[f]{the set of slicing criteria}}\\
    \If{F is not called}{
        \GenCallFunc{F} \textcolor[HTML]{70ad48}{ \tcp*[f]{generate call function}} \\
    }
    $S_{var} \leftarrow$ \GetFuncVar{F} \textcolor[HTML]{70ad48}{ \tcp*[f]{get function variables}} \\
    \For{$v \in S_{var}$}{ 
        $sc \leftarrow$ \InsSliCrit{F, v} \textcolor[HTML]{70ad48}{ \tcp*[f]{construct and insert slicing criteria}} \\
        $S_{sc}$.\append($sc$) \\
    }
    \For{$s \in S_{sc}$}{
        $fs \leftarrow$ \Slicing{F, s} \textcolor[HTML]{70ad48}{ \tcp*[f]{generate slice}} \\
        \If{$fs$ no empty}{
            $S_{fs}$.\append($fs$)\\   
        }
    }
    $S_{fs} \leftarrow$ \SimplifySlicing{$S_{fs}$} \textcolor[HTML]{70ad48}{ \tcp*[f]{simplify slices}} \\
    \Return $S_{fs}$
}
\end{algorithm}

To address the above challenges, we propose an automatic function slicing algorithm. Specifically, as shown in Algorithm~\ref{algo: Automatic Function Slicing}, the algorithm takes the function $F$ to be sliced as input. First, it checks whether $F$ is invoked by any caller. If not, a simple wrapper function is created for $F$ (lines~4–5) to satisfy the prerequisite for performing program slicing. Next, the algorithm extracts key variables from $F$ (line~6). In this work, we select local variables defined in the function, excluding those defined within loops. Variables defined in loops mainly serve two purposes: (1) recording the current iteration state, and (2) storing temporary results. The slices generated from such variables are typically subsumed by slices generated from variables defined outside the loops. Excluding these variables therefore improves slicing efficiency. Then, each variable in the extracted set is used as a parameter of a synthetic function, and the corresponding statement is inserted at the end of the function as a slicing criterion (lines~7–9). Subsequently, the algorithm iterates over the specified slicing criteria and generates function slices (lines~10–13). However, we observe that the resulting slices may overlap. To address this issue, we apply a simplification algorithm to reduce the slice set (line~14), thereby improving the efficiency of subsequent specification generation.

The simplification algorithm reduces the input slice set $S$ using a greedy strategy. Specifically, as shown in Algorithm~\ref{algo: Simplified Slice Set}, the algorithm first counts all code statements involved in the slices and stores them in the set $S_{as}$ (lines~5–6). It then repeatedly selects from $S$ the function slice that covers the largest number of previously uncovered statements (lines~10–14). The selected slice is added to the result set $S_{ss}$, while the set of covered statements $S_{cs}$ is updated and the slice is removed from $S$ (lines~15–17). This process continues until all statements are covered (line~7).

\begin{algorithm}[tbp!]
\scriptsize
\caption{Simplified Slice Set}
\label{algo: Simplified Slice Set} 
\SetKwProg{Fn}{Function}{\string:}{}
\SetKwInOut{Input}{input}
\SetKwInOut{Output}{output}
\SetKwFunction{append}{append}
\SetKwFunction{SimplifySlicing}{SimplifySlicing}
\SetKwFunction{update}{update}
\SetKwFunction{len}{len}
\SetKwFunction{remove}{remove}
\Input{$S$: the set of slices} 
\Output{$S_{ss}$: the set of selected slices}
\BlankLine 

\Fn{\SimplifySlicing{$S$}}{
    $S_{as} = \emptyset$ \textcolor[HTML]{70ad48}{ \tcp*[f]{the set of all statements}} \\
    $S_{cs} = \emptyset$ \textcolor[HTML]{70ad48}{ \tcp*[f]{the set of covered statements}} \\ 
    $S_{ss} = \emptyset$ \textcolor[HTML]{70ad48}{ \tcp*[f]{the set of selected slices}} \\
    \For{$s \in S$}{
        $S_{as}$.\update{s} \textcolor[HTML]{70ad48}{ \tcp*[f]{get all slices' statements}} \\
    }
    \While{$S_{cs} \not = S_{as}$}{
        max\_uc = -1 \textcolor[HTML]{70ad48}{ \tcp*[f]{the maximum number of uncovered statements}} \\
        best\_slice = $\emptyset$ \\
        \For{$s \in S$}{
            uc\_stat = s - $S_{cs}$ \textcolor[HTML]{70ad48}{ \tcp*[f]{get the uncovered statements in slice}} \\
            \If(\textcolor[HTML]{70ad48}{ \tcp*[f]{find the slice that covers more statements}}){\len{uc\_stat} \textgreater max\_uc}{
                max\_uc = \len{uc\_stat} \\
                best\_slice = s \\
            }
        }
        $S_{ss}$.\append{best\_slice} \\
        $S_{cs}$.\update{best\_slice} \\
        $S$.\remove{best\_slice} \\
    }
    \Return $S_{ss}$
}
\end{algorithm}

\subsection{Specification Generation}
\label{subsec: Specification Generation}
In this phase, SLD-Spec feeds the function slices into LLMs for specification generation. If a slice contains substructures, specifications are generated for these substructures with priority. Similar to existing approaches~\cite{Ma-01,Wen-01}, our prompt consists of three parts: role setting, few-shot examples, and the slice code. The role setting defines the role that the LLM should assume in the dialogue, allowing them to generate more contextually appropriate responses based on the user's request. The inclusion of few-shot examples helps the LLM learn from these instances and generate the desired output format for subsequent processing.

\subsection{Logical Deletion}
\label{subsec: Logical Deletion}
To mitigate the issues of correct specifications being discarded due to insufficient context, as well as the unreliability of filtering specifications solely using verification tools, we introduce a logical deletion phase between the generation and verification phases. Specifically, we first summarize seven types of specifications that are prone to the above issues. Next, we propose the concept of logical deletion, which leverages LLMs to perform non-strict code reasoning to assess the correctness of specifications. Finally, when constructing the prompt for logical deletion, we instantiate common erroneous specification types as few-shot examples and include them alongside correct specifications, thereby enhancing the LLMs’ judgment capability.

Table~\ref{table: Common Erroneous Specification Types} presents seven common erroneous specification types, along with representative specification examples and corresponding explanations based on the source code shown in Fig.~\ref{pic: motivation}. In addition, the causes of these seven types of errors are detailed as follows:
\begin{itemize}
    \item \textbf{Incorrect Boundary Condition}: Although LLMs are adept at capturing the overall semantics of programs, they often misinterpret fine-grained relationships, such as variable scope.
    \item \textbf{Incorrect Pattern Summarization}: Since existing LLMs are not trained on specialized specification datasets, they may fail to accurately interpret programs and generate the corresponding specifications.
    \item \textbf{Invariant Misalignment}: Influenced by unrelated code segments, LLMs may directly apply irrelevant specifications.
    \item \textbf{False Modifiability}: LLMs fail to understand the semantics of {\consolaslike{loop assigns}} and instead directly apply loop variables to this type of specification without proper analysis.
    \item \textbf{Assigns Misalignment}: Similar to invariant misalignment, LLMs may inappropriately apply the specifications of unrelated loops.
    \item \textbf{Uniqueness Conflict}: LLMs may produce multiple {\consolaslike{loop variant}} statements, thereby violating the uniqueness requirement of this specification type.
    \item \textbf{Invalid Definition}: LLMs misconstrue the meaning of {\consolaslike{loop variant}} and instead apply variables from the loop condition to examples that appear in the prompt.
\end{itemize}

\begin{table*}[tbp!]
\caption{Common Erroneous Specification Types}
\label{table: Common Erroneous Specification Types}
\adjustbox{max width=\textwidth}{
\begin{tabular}{@{}c|c|l|l@{}}
\toprule
Type of Specification & Error Type                           & \multicolumn{1}{c|}{Example}         & \multicolumn{1}{c}{Explanation of Error} \\ \midrule
\multirow{6}{*}{loop invariant} & Incorrect Boundary Condition         & loop invariant c \textless~x;        &  the variable c is equal to x after the loop has fully executed.                                \\ \cmidrule(l){2-4} 
                           & Incorrect Pattern Summarization      & loop invariant *prod\_res == x * y;  &  \makecell[l]{the value of the *prod\_res is equal to the number of loop \\ executions x - c multiplied by the increment y.}                               \\ \cmidrule(l){2-4} 
                           & Invariant Misalignment               & loop invariant *x\_res == x - a;     &  \makecell[l]{the object *x\_res described by this specification does not \\ exist in the loop 2.}                               \\ \midrule
\multirow{3}{*}{loop assigns}   & False Modifiability                  & loop assigns y;                      &  the variable y is not modified in the loop.                               \\ \cmidrule(l){2-4} 
                           & Assigns Misalignment                 & loop assigns *x\_res;                &  \makecell[l]{the object *x\_res described by this specification does not \\ exist in the loop 2.}                               \\ \midrule
\multirow{4}{*}{loop variant}   & \multirow{2}{*}{Uniqueness Conflict} & loop variant c;                      &  \multirow{2}{*}{\makecell[l]{multiple loop variant statements appearing simultaneously, \\ defined as syntax error.}}               \\
                           &                                      & loop variant x - c;                  &                                 \\ \cmidrule(l){2-4} 
                           & Invalid Definition                   & loop variant x - c;                  &  \makecell[l]{the expression x - c does not satisfy the requirement \\ of decrease for loop variant.}                               \\ \bottomrule
\end{tabular}
}
\end{table*}

We introduce a logical deletion stage prior to specification verification, with the goal of leveraging LLMs to assess the plausibility and potential satisfiability of specifications, without requiring them to be formally provable in the given context. However, since existing LLMs have not been trained on specialized specification datasets, directly providing LLMs with all candidate specifications and the corresponding program source code and requesting judgment outputs often results in low accuracy. This approach may also lead to inconsistencies between the number of generated outputs and the number of specifications under verification.

Since LLMs have been extensively trained on natural language text and code data~\cite{Chen-01,Hou-01}, we draw on the idea of chain-of-thought reasoning~\cite{Wang-01} and propose a logical deletion method consisting of four steps: exclusion, comprehension, reasoning, and output. First, LLMs exclude specifications that are irrelevant to the current code based on the occurrence of variables in the specifications. This step filters out unrelated specifications in advance, thereby reducing the workload of subsequent steps. Second, LLMs interpret the candidate specifications according to different specification types and describe them in informal natural language. Third, we treat these interpretations as program design requirements and feed them, together with the source program, into LLMs, requiring the models to reason about whether the program satisfies each requirement based on their understanding of the source code. Finally, LLMs output the reasoning results—true or false—according to the affirmative or negative stance reflected in the generated explanations.

Additionally, when constructing prompts for logical deletion, we randomly select common specification error types and generate corresponding erroneous specifications for the program examples. This allows LLMs to reference these error types during the explanation process, thereby mitigating hallucinations.

\subsection{Specification Verification}
\label{subsec: Specification Verification}
Following the three phases described above, SLD-Spec generates candidate specifications for each slice. Unlike other approaches, SLD-Spec does not immediately verify the candidate specifications for individual slices. Instead, it delays verification until specifications for all slices belonging to the same function have been generated, at which point verification is performed collectively. There are two primary reasons for this strategy. First, different slices may produce identical specifications, and verifying each slice individually would be highly time-consuming. Second, a specification that is correct for an individual slice may not hold for the entire function. For example, if a slice modifies no variables, the {\consolaslike{assigns nothing}} statement is correct for that slice. However, if other slices of the same function modify variables, this specification becomes incorrect for the function as a whole. Therefore, before initiating verification, SLD-Spec integrates all slice-level specifications and removes any that conflict with one another.

SLD-Spec employs the Frama-C verification tool~\cite{Kirchner-01} to verify the integrated specifications. If the verification tool reports a failure, SLD-Spec identifies and locates the incorrect specifications, removes them, and then re-verifies the remaining set. This process is repeated until all specifications are either removed or successfully verified. If no verification failures occur, SLD-Spec outputs the final set of retained specifications.

\section{Evaluation}
Our experiments are designed to address the following research questions (RQs): 

\textbf{RQ1: How does SLD-Spec perform compared to existing approaches on datasets containing simple programs?} Existing LLM-based specification generation approaches have demonstrated strong performance on relatively simple programs. Evaluating SLD-Spec under this setting helps determine whether the proposed method maintains competitive performance.

\textbf{RQ2: How does SLD-Spec perform compared to existing approaches on datasets containing complex control flow?} Programs with complex control-flow structures pose significant challenges for specification generation. This RQ investigates how SLD-Spec behaves in such scenarios and whether its design is better suited to handling increased program complexity.

\textbf{RQ3: What is the impact of program slicing and logical deletion on the performance of SLD-Spec?} Program slicing and logical deletion are the core components of SLD-Spec. Understanding their individual and combined impact through an ablation study is essential for assessing the effectiveness of these design choices and identifying their respective contributions to overall performance.

After discussing the above RQs, we further selected four representative cases that failed verification in the experiments and conducted a detailed analysis of their failure causes. The results of this analysis are presented in Appendix~\ref{Case Studies of Verification Failure}.

\subsection{Experimental Setup}
In the program slicing phase, we adopt DG~\cite{Chalupa-01} as the slicing tool in our experiments. DG constructs a program dependence graph through control dependence analysis, thereby supporting static backward program slicing. In the generation phase, SLD-Spec communicates with LLMs via the OpenAI API~\cite{OpenAI-01}. In addition, to ensure experimental rigor, we set the \textit{temperature} parameter to 0.7 and the \textit{max\_tokens} parameter to 2048. In the verification phase, we choose Frama-C as the program verification tool. Frama-C verifies the correctness of ACSL specifications using the weakest precondition plugin~\cite{Patrick-02}, and we set the specification verification timeout to 8 seconds. All experiments are conducted on a workstation equipped with an Intel Core i9-13900K CPU, 64 GB of memory, an NVIDIA GeForce RTX 4080 GPU, and running Ubuntu 22.04 LTS.

\subsubsection{Baselines.}
For RQ1 and RQ2, we compare our approach with two SOTA LLM-based specification generation methods: AutoSpec~\cite{Wen-01} and SpecGen~\cite{Ma-01}. AutoSpec combines LLMs with static analysis and guides the attention of LLMs through a hierarchical generation strategy, thereby producing accurate specifications. SpecGen, in contrast, adopts a conversation-driven approach to progressively improve the accuracy of generated specifications. In addition, for programs that fail verification, SpecGen employs a heuristic mutated-specification selection algorithm to reduce the size of the specification set to be verified.

\subsubsection{Dataset.}
To accurately evaluate the performance of different approaches, we conduct experiments on two datasets. The \textit{frama-c-problems} dataset~\cite{Manav-01} is derived from prior work~\cite{Wen-01}, in which 51 simple programs are categorized into eight classes. However, most functions in this benchmark perform relatively simple tasks, and the majority of programs contain at most one control-flow construct, making it insufficient for representing complex programs. To address this limitation, we construct a new dataset, \textit{complex-func}, based on \textit{frama-c-problems}. The new dataset is obtained by renaming variables, composing multiple programs, and transforming code locations in the original programs. The \textit{complex-func} dataset can be divided into the following four categories:
\begin{itemize}
    \item \textbf{Parallel Single-Path Loop}: Programs in this category contain multiple parallel single-path loops, each consisting of a single loop construct without any branch statements in the loop body.
    \item \textbf{Single Multi-Path Loop}: Programs in this category contain a single multi-path loop whose body includes multiple branch statements.
    \item \textbf{Conditional-Enhanced Single-Path Loop}: Programs in this category are based on the \textit{Parallel Single-Path Loop} structure but introduce an additional branch statement at varying positions in the program.
    \item \textbf{Nested Loop}: Programs in this category contain a nested loop structure in which each inner loop represents a distinct task.
\end{itemize}

\subsubsection{Evaluation Metrics.}
Building on prior work~\cite{Ma-01,Wen-01}, we introduce more fine-grained evaluation metrics. To mitigate the impact of randomness in LLM outputs on the experimental results, we conduct five independent runs for each program. All reported metrics—except for \textit{NPP}, which inherently requires multiple runs—are computed as averages over these repeated experiments.
\begin{itemize}
    \item \textbf{Proportion of Correct and Relevant Specifications After Verification (PCRSAV)}: Although the generated specifications may be syntactically and semantically correct, they can still include irrelevant specifications that are meaningless in practice and ultimately require manual removal. Therefore, a higher value of this metric indicates a stronger understanding capability of the generation approach.
    \item \textbf{Number of Assertions Passed Verification (NAV)}: Assertions are used to help prove whether the function behavior described by the generated specifications satisfies the required properties. Even if a function fails overall verification, a higher value of this metric indicates broader coverage of program behaviors by the specifications.
    \item \textbf{Number of Program Verification Passes (NPP)}: Due to the inherent randomness of LLM-generated content, the specifications produced in different runs are not always complete. Therefore, we use this metric to evaluate the robustness of different approaches across multiple runs.
    \item \textbf{Running Time (RT)}: We use this metric to evaluate the efficiency of different approaches. It measures the total time from receiving the input program, through specification generation and verification, to producing the final specifications.
\end{itemize}

\subsection{Experimental Results}
\subsubsection{RQ1 Effectiveness of frama-c-problems.}
\begin{table}
\scriptsize
\centering
\caption{Effectiveness of Different Approaches on the frama-c-problems}
\label{table: Effectiveness of Different Approaches on the frama-c-problems}
\begin{threeparttable}
\begin{tabular}{@{}c|lll|ll|ll|ll@{}}
\toprule
\multicolumn{4}{c|}{\textbf{Dataset Information}}     & \multicolumn{2}{c|}{\textbf{AutoSpec}}    & \multicolumn{2}{c|}{\textbf{SpecGen}}    & \multicolumn{2}{c}{\textbf{SLD-Spec}}                                                                                                        \\ \midrule
\multicolumn{1}{c|}{\textbf{Type}}                   & \multicolumn{1}{c|}{\textbf{Program}}                   & \multicolumn{1}{c|}{\textbf{LoC}}   & \multicolumn{1}{c|}{\textbf{NoL}}  & \multicolumn{1}{c}{\textbf{NPP}} & \multicolumn{1}{c|}{\textbf{RT(s)}}  & \multicolumn{1}{c}{\textbf{NPP}} & \multicolumn{1}{c|}{\textbf{RT(s)}}  & \multicolumn{1}{c}{\textbf{NPP}} & \multicolumn{1}{c}{\textbf{RT(s)}} \\ \midrule

\multicolumn{1}{c|}{\multirow{12}{*}{general\_wp\_problems}}
& \multicolumn{1}{l|}{absolute\_value.c}         & \multicolumn{1}{c|}{13}    &  \multicolumn{1}{c|}{0}      & \multicolumn{1}{c}{5/5}    & \multicolumn{1}{c|}{54.00}     & \multicolumn{1}{c}{5/5}    & \multicolumn{1}{c|}{3.56}     & \multicolumn{1}{c}{5/5}    & \multicolumn{1}{c}{12.46}                        \\
& \multicolumn{1}{l|}{add.c}                     & \multicolumn{1}{c|}{14}    &  \multicolumn{1}{c|}{0}      & \multicolumn{1}{c}{5/5}    & \multicolumn{1}{c|}{21.74}     & \multicolumn{1}{c}{5/5}    & \multicolumn{1}{c|}{2.27}     & \multicolumn{1}{c}{5/5}    & \multicolumn{1}{c}{4.00}                        \\
& \multicolumn{1}{l|}{ani.c}                     & \multicolumn{1}{c|}{15}    &  \multicolumn{1}{c|}{1}      & \multicolumn{1}{c}{0/5}    & \multicolumn{1}{c|}{65.05}       & \multicolumn{1}{c}{3/5}    & \multicolumn{1}{c|}{51.10}     & \multicolumn{1}{c}{3/5}    & \multicolumn{1}{c}{67.32}                        \\
& \multicolumn{1}{l|}{diff.c}                    & \multicolumn{1}{c|}{8}     &  \multicolumn{1}{c|}{0}      & \multicolumn{1}{c}{5/5}    & \multicolumn{1}{c|}{2.85}       & \multicolumn{1}{c}{5/5}    & \multicolumn{1}{c|}{2.06}     & \multicolumn{1}{c}{5/5}    & \multicolumn{1}{c}{2.80}                        \\
& \multicolumn{1}{l|}{gcd.c}                     & \multicolumn{1}{c|}{17}    &  \multicolumn{1}{c|}{0}      & \multicolumn{1}{c}{0/5}    & \multicolumn{1}{c|}{188.57}       & \multicolumn{1}{c}{0/5}    & \multicolumn{1}{c|}{164.17}     & \multicolumn{1}{c}{0/5}    & \multicolumn{1}{c}{23.09}                        \\
& \multicolumn{1}{l|}{max\_of\_2.c}              & \multicolumn{1}{c|}{12}    &  \multicolumn{1}{c|}{0}      & \multicolumn{1}{c}{5/5}    & \multicolumn{1}{c|}{5.69}       & \multicolumn{1}{c}{5/5}    & \multicolumn{1}{c|}{2.71}     & \multicolumn{1}{c}{5/5}    & \multicolumn{1}{c}{2.63}                        \\
& \multicolumn{1}{l|}{power.c}                   & \multicolumn{1}{c|}{17}    &  \multicolumn{1}{c|}{1}      & \multicolumn{1}{c}{0/5}    & \multicolumn{1}{c|}{82.2}       & \multicolumn{1}{c}{0/5}    & \multicolumn{1}{c|}{94.76}     & \multicolumn{1}{c}{0/5}    & \multicolumn{1}{c}{35.97}                        \\
& \multicolumn{1}{l|}{simple\_interest.c}        & \multicolumn{1}{c|}{10}    &  \multicolumn{1}{c|}{0}      & \multicolumn{1}{c}{5/5}    & \multicolumn{1}{c|}{4.04}       & \multicolumn{1}{c}{5/5}    & \multicolumn{1}{c|}{2.69}     & \multicolumn{1}{c}{5/5}    & \multicolumn{1}{c}{3.66}                        \\
& \multicolumn{1}{l|}{swap.c}                    & \multicolumn{1}{c|}{12}    &  \multicolumn{1}{c|}{0}      & \multicolumn{1}{c}{5/5}    & \multicolumn{1}{c|}{8.43}       & \multicolumn{1}{c}{5/5}    & \multicolumn{1}{c|}{4.20}     & \multicolumn{1}{c}{5/5}    & \multicolumn{1}{c}{4.59}                        \\
& \multicolumn{1}{l|}{triangle\_angles.c}        & \multicolumn{1}{c|}{13}    &  \multicolumn{1}{c|}{0}      & \multicolumn{1}{c}{5/5}    & \multicolumn{1}{c|}{9.89}       & \multicolumn{1}{c}{5/5}    & \multicolumn{1}{c|}{7.79}     & \multicolumn{1}{c}{5/5}    & \multicolumn{1}{c}{5.10}                        \\
& \multicolumn{1}{l|}{triangle\_sides.c}         & \multicolumn{1}{c|}{14}    &  \multicolumn{1}{c|}{0}      & \multicolumn{1}{c}{5/5}    & \multicolumn{1}{c|}{3.98}       & \multicolumn{1}{c}{5/5}    & \multicolumn{1}{c|}{4.31}     & \multicolumn{1}{c}{5/5}    & \multicolumn{1}{c}{6.66}                        \\
& \multicolumn{1}{l|}{wp1.c}                     & \multicolumn{1}{c|}{13}    &  \multicolumn{1}{c|}{0}      & \multicolumn{1}{c}{0/5}    & \multicolumn{1}{c|}{59.08}       & \multicolumn{1}{c}{0/5}    & \multicolumn{1}{c|}{79.09}     & \multicolumn{1}{c}{0/5}    & \multicolumn{1}{c}{5.16}                        \\ \midrule

\multicolumn{1}{c|}{\multirow{8}{*}{pointers}}
& \multicolumn{1}{l|}{add\_pointers.c}           & \multicolumn{1}{c|}{15}    &  \multicolumn{1}{c|}{0}      & \multicolumn{1}{c}{5/5}    & \multicolumn{1}{c|}{25.01}       & \multicolumn{1}{c}{5/5}    & \multicolumn{1}{c|}{3.05}     & \multicolumn{1}{c}{5/5}    & \multicolumn{1}{c}{3.52}                        \\
& \multicolumn{1}{l|}{add\_pointers\_3\_vars.c}  & \multicolumn{1}{c|}{16}    &  \multicolumn{1}{c|}{0}      & \multicolumn{1}{c}{5/5}    & \multicolumn{1}{c|}{12.13}       & \multicolumn{1}{c}{5/5}    & \multicolumn{1}{c|}{2.34}     & \multicolumn{1}{c}{5/5}    & \multicolumn{1}{c}{3.82}                        \\
& \multicolumn{1}{l|}{div\_rem.c}                & \multicolumn{1}{c|}{11}    &  \multicolumn{1}{c|}{0}      & \multicolumn{1}{c}{5/5}    & \multicolumn{1}{c|}{3.90}       & \multicolumn{1}{c}{0/5}    & \multicolumn{1}{c|}{86.47}     & \multicolumn{1}{c}{5/5}    & \multicolumn{1}{c}{11.88}                        \\
& \multicolumn{1}{l|}{incr\_a\_by\_b.c}          & \multicolumn{1}{c|}{12}    &  \multicolumn{1}{c|}{0}      & \multicolumn{1}{c}{5/5}    & \multicolumn{1}{c|}{5.71}       & \multicolumn{1}{c}{0/5}    & \multicolumn{1}{c|}{103.24}     & \multicolumn{1}{c}{5/5}    & \multicolumn{1}{c}{7.39}                        \\
& \multicolumn{1}{l|}{max\_pointers.c}           & \multicolumn{1}{c|}{12}    &  \multicolumn{1}{c|}{0}      & \multicolumn{1}{c}{5/5}    & \multicolumn{1}{c|}{11.41}       & \multicolumn{1}{c}{5/5}    & \multicolumn{1}{c|}{3.86}     & \multicolumn{1}{c}{5/5}    & \multicolumn{1}{c}{3.86}                        \\
& \multicolumn{1}{l|}{order\_3.c}                & \multicolumn{1}{c|}{31}    &  \multicolumn{1}{c|}{0}      & \multicolumn{1}{c}{0/5}    & \multicolumn{1}{c|}{149.18}       & \multicolumn{1}{c}{0/5}    & \multicolumn{1}{c|}{43.30}     & \multicolumn{1}{c}{0/5}    & \multicolumn{1}{c}{19.69}                        \\ 
& \multicolumn{1}{l|}{reset\_1st.c}              & \multicolumn{1}{c|}{14}    &  \multicolumn{1}{c|}{0}      & \multicolumn{1}{c}{3/5}    & \multicolumn{1}{c|}{37.87}       & \multicolumn{1}{c}{1/5}    & \multicolumn{1}{c|}{97.71}     & \multicolumn{1}{c}{5/5}    & \multicolumn{1}{c}{7.45}                        \\
& \multicolumn{1}{l|}{swap.c}                    & \multicolumn{1}{c|}{12}    &  \multicolumn{1}{c|}{0}      & \multicolumn{1}{c}{5/5}    & \multicolumn{1}{c|}{3.72}       & \multicolumn{1}{c}{5/5}    & \multicolumn{1}{c|}{4.45}     & \multicolumn{1}{c}{5/5}    & \multicolumn{1}{c}{6.43}                        \\ \midrule

\multicolumn{1}{c|}{\multirow{8}{*}{loops}}
& \multicolumn{1}{l|}{1.c}                       & \multicolumn{1}{c|}{8}     &  \multicolumn{1}{c|}{1}      & \multicolumn{1}{c}{5/5}    & \multicolumn{1}{c|}{6.35}       & \multicolumn{1}{c}{5/5}    & \multicolumn{1}{c|}{2.15}     & \multicolumn{1}{c}{5/5}    & \multicolumn{1}{c}{14.58}                        \\
& \multicolumn{1}{l|}{2.c}                       & \multicolumn{1}{c|}{14}    &  \multicolumn{1}{c|}{1}      & \multicolumn{1}{c}{0/5}    & \multicolumn{1}{c|}{211.27}       & \multicolumn{1}{c}{0/5}    & \multicolumn{1}{c|}{35.51}     & \multicolumn{1}{c}{0/5}    & \multicolumn{1}{c}{33.62}                         \\
& \multicolumn{1}{l|}{3.c}                       & \multicolumn{1}{c|}{15}    &  \multicolumn{1}{c|}{1}      & \multicolumn{1}{c}{0/5}    & \multicolumn{1}{c|}{46.81}       & \multicolumn{1}{c}{4/5}    & \multicolumn{1}{c|}{49.85}     & \multicolumn{1}{c}{4/5}    & \multicolumn{1}{c}{21.03}                         \\
& \multicolumn{1}{l|}{4.c}                       & \multicolumn{1}{c|}{15}    &  \multicolumn{1}{c|}{1}      & \multicolumn{1}{c}{0/5}    & \multicolumn{1}{c|}{207.26}       & \multicolumn{1}{c}{0/5}    & \multicolumn{1}{c|}{30.15}     & \multicolumn{1}{c}{0/5}    & \multicolumn{1}{c}{32.11}                         \\
& \multicolumn{1}{l|}{fact.c}                    & \multicolumn{1}{c|}{15}    &  \multicolumn{1}{c|}{1}      & \multicolumn{1}{c}{0/5}    & \multicolumn{1}{c|}{111.20}       & \multicolumn{1}{c}{0/5}    & \multicolumn{1}{c|}{22.41}     & \multicolumn{1}{c}{0/5}    & \multicolumn{1}{c}{34.40}                         \\
& \multicolumn{1}{l|}{mult.c}                    & \multicolumn{1}{c|}{13}    &  \multicolumn{1}{c|}{1}      & \multicolumn{1}{c}{0/5}    & \multicolumn{1}{c|}{42.38}       & \multicolumn{1}{c}{5/5}    & \multicolumn{1}{c|}{3.03}     & \multicolumn{1}{c}{4/5}    & \multicolumn{1}{c}{23.00}                         \\
& \multicolumn{1}{l|}{sum\_digits.c}             & \multicolumn{1}{c|}{15}    &  \multicolumn{1}{c|}{1}      & \multicolumn{1}{c}{0/5}    & \multicolumn{1}{c|}{71.73}       & \multicolumn{1}{c}{0/5}    & \multicolumn{1}{c|}{34.67}     & \multicolumn{1}{c}{0/5}    & \multicolumn{1}{c}{30.67}                         \\
& \multicolumn{1}{l|}{sum\_even.c}               & \multicolumn{1}{c|}{14}    &  \multicolumn{1}{c|}{1}      & \multicolumn{1}{c}{0/5}    & \multicolumn{1}{c|}{81.15}       & \multicolumn{1}{c}{4/5}    & \multicolumn{1}{c|}{39.87}     & \multicolumn{1}{c}{5/5}    & \multicolumn{1}{c}{47.34}                         \\ \midrule

\multicolumn{1}{c|}{\multirow{8}{*}{immutable\_arrays}}
& \multicolumn{1}{l|}{array\_sum.c}              & \multicolumn{1}{c|}{14}    &  \multicolumn{1}{c|}{1}      & \multicolumn{1}{c}{0/5}    & \multicolumn{1}{c|}{79.95}       & \multicolumn{1}{c}{0/5}    & \multicolumn{1}{c|}{44.42}     & \multicolumn{1}{c}{0/5}    & \multicolumn{1}{c}{20.57}                         \\
& \multicolumn{1}{l|}{binary\_search.c}          & \multicolumn{1}{c|}{21}    &  \multicolumn{1}{c|}{1}      & \multicolumn{1}{c}{0/5}    & \multicolumn{1}{c|}{1320.39}       & \multicolumn{1}{c}{0/5}    & \multicolumn{1}{c|}{253.49}     & \multicolumn{1}{c}{0/5}    & \multicolumn{1}{c}{60.57}                         \\ 
& \multicolumn{1}{l|}{check\_evens\_in\_array.c} & \multicolumn{1}{c|}{16}    &  \multicolumn{1}{c|}{1}      & \multicolumn{1}{c}{0/5}    & \multicolumn{1}{c|}{22.74}       & \multicolumn{1}{c}{5/5}    & \multicolumn{1}{c|}{6.16}     & \multicolumn{1}{c}{4/5}    & \multicolumn{1}{c}{23.52}                         \\
& \multicolumn{1}{l|}{max.c}                     & \multicolumn{1}{c|}{22}    &  \multicolumn{1}{c|}{1}      & \multicolumn{1}{c}{5/5}    & \multicolumn{1}{c|}{45.04}       & \multicolumn{1}{c}{5/5}    & \multicolumn{1}{c|}{11.06}     & \multicolumn{1}{c}{5/5}    & \multicolumn{1}{c}{38.28}                         \\
& \multicolumn{1}{l|}{occurences\_of\_x.c}       & \multicolumn{1}{c|}{23}    &  \multicolumn{1}{c|}{1}      & \multicolumn{1}{c}{4/5}    & \multicolumn{1}{c|}{39.42}       & \multicolumn{1}{c}{1/5}    & \multicolumn{1}{c|}{46.18}     & \multicolumn{1}{c}{0/5}    & \multicolumn{1}{c}{98.31}                         \\ 
& \multicolumn{1}{l|}{sample.c}                  & \multicolumn{1}{c|}{13}    &  \multicolumn{1}{c|}{1}      & \multicolumn{1}{c}{0/5}    & \multicolumn{1}{c|}{120.10}       & \multicolumn{1}{c}{0/5}    & \multicolumn{1}{c|}{128.67}     & \multicolumn{1}{c}{5/5}    & \multicolumn{1}{c}{66.65}                         \\ 
& \multicolumn{1}{l|}{search.c}                  & \multicolumn{1}{c|}{15}    &  \multicolumn{1}{c|}{1}      & \multicolumn{1}{c}{0/5}    & \multicolumn{1}{c|}{26.51}       & \multicolumn{1}{c}{5/5}    & \multicolumn{1}{c|}{5.09}     & \multicolumn{1}{c}{3/5}    & \multicolumn{1}{c}{28.64}                         \\
& \multicolumn{1}{l|}{search\_2.c}               & \multicolumn{1}{c|}{16}    &  \multicolumn{1}{c|}{1}      & \multicolumn{1}{c}{0/5}    & \multicolumn{1}{c|}{141.4}       & \multicolumn{1}{c}{5/5}    & \multicolumn{1}{c|}{4.19}     & \multicolumn{1}{c}{5/5}    & \multicolumn{1}{c}{38.36}                         \\ \midrule

\multicolumn{1}{c|}{\multirow{2}{*}{mutable\_arrays}}
& \multicolumn{1}{l|}{array\_double.c}           & \multicolumn{1}{c|}{17}    &  \multicolumn{1}{c|}{1}      & \multicolumn{1}{c}{0/5}    & \multicolumn{1}{c|}{122.46}       & \multicolumn{1}{c}{5/5}    & \multicolumn{1}{c|}{9.45}     & \multicolumn{1}{c}{0/5}    & \multicolumn{1}{c}{25.81}                         \\
& \multicolumn{1}{l|}{bubble\_sort.c}            & \multicolumn{1}{c|}{20}    &  \multicolumn{1}{c|}{2}      & \multicolumn{1}{c}{0/5}    & \multicolumn{1}{c|}{1009.09}       & \multicolumn{1}{c}{0/5}    & \multicolumn{1}{c|}{233.66}     & \multicolumn{1}{c}{0/5}    & \multicolumn{1}{c}{167.08}                         \\ \midrule

\multicolumn{1}{c|}{\multirow{3}{*}{more\_arrays}}
& \multicolumn{1}{l|}{equal\_arrays.c}           & \multicolumn{1}{c|}{14}    &  \multicolumn{1}{c|}{1}      & \multicolumn{1}{c}{2/5}    & \multicolumn{1}{c|}{28.78}       & \multicolumn{1}{c}{5/5}    & \multicolumn{1}{c|}{4.41}     & \multicolumn{1}{c}{2/5}    & \multicolumn{1}{c}{34.74}                         \\
& \multicolumn{1}{l|}{replace\_evens.c}          & \multicolumn{1}{c|}{14}    &  \multicolumn{1}{c|}{1}      & \multicolumn{1}{c}{2/5}    & \multicolumn{1}{c|}{32.58}       & \multicolumn{1}{c}{4/5}    & \multicolumn{1}{c|}{45.09}     & \multicolumn{1}{c}{5/5}    & \multicolumn{1}{c}{19.79}                         \\
& \multicolumn{1}{l|}{reverse\_array.c}          & \multicolumn{1}{c|}{17}    &  \multicolumn{1}{c|}{1}      & \multicolumn{1}{c}{0/5}    & \multicolumn{1}{c|}{131.98}       & \multicolumn{1}{c}{3/5}    & \multicolumn{1}{c|}{94.91}     & \multicolumn{1}{c}{4/5}    & \multicolumn{1}{c}{86.17}                         \\ \midrule

\multicolumn{1}{c|}{\multirow{5}{*}{arrays\_and\_loops}}
& \multicolumn{1}{l|}{1.c}                       & \multicolumn{1}{c|}{9}     &  \multicolumn{1}{c|}{0}      & \multicolumn{1}{c}{5/5}    & \multicolumn{1}{c|}{2.97}       & \multicolumn{1}{c}{5/5}    & \multicolumn{1}{c|}{3.44}     & \multicolumn{1}{c}{5/5}    & \multicolumn{1}{c}{2.65}                         \\
& \multicolumn{1}{l|}{2.c}                       & \multicolumn{1}{c|}{19}    &  \multicolumn{1}{c|}{1}      & \multicolumn{1}{c}{5/5}    & \multicolumn{1}{c|}{47.31}       & \multicolumn{1}{c}{5/5}    & \multicolumn{1}{c|}{4.28}     & \multicolumn{1}{c}{5/5}    & \multicolumn{1}{c}{42.01}                         \\
& \multicolumn{1}{l|}{3.c}                       & \multicolumn{1}{c|}{18}    &  \multicolumn{1}{c|}{0}      & \multicolumn{1}{c}{5/5}    & \multicolumn{1}{c|}{12.45}       & \multicolumn{1}{c}{5/5}    & \multicolumn{1}{c|}{22.82}     & \multicolumn{1}{c}{5/5}    & \multicolumn{1}{c}{8.39}                         \\
& \multicolumn{1}{l|}{4.c}                       & \multicolumn{1}{c|}{15}    &  \multicolumn{1}{c|}{1}      & \multicolumn{1}{c}{0/5}    & \multicolumn{1}{c|}{40.87}       & \multicolumn{1}{c}{4/5}    & \multicolumn{1}{c|}{43.60}     & \multicolumn{1}{c}{5/5}    & \multicolumn{1}{c}{21.83}                         \\
& \multicolumn{1}{l|}{5.c}                       & \multicolumn{1}{c|}{16}    &  \multicolumn{1}{c|}{1}      & \multicolumn{1}{c}{0/5}    & \multicolumn{1}{c|}{136.31}       & \multicolumn{1}{c}{4/5}    & \multicolumn{1}{c|}{63.04}     & \multicolumn{1}{c}{3/5}    & \multicolumn{1}{c}{76.83}                         \\ \midrule

\multicolumn{1}{c|}{\multirow{5}{*}{miscellaneous}}
& \multicolumn{1}{l|}{array\_find.c}             & \multicolumn{1}{c|}{17}    &  \multicolumn{1}{c|}{1}      & \multicolumn{1}{c}{2/5}    & \multicolumn{1}{c|}{54.24}       & \multicolumn{1}{c}{0/5}    & \multicolumn{1}{c|}{86.37}     & \multicolumn{1}{c}{2/5}    & \multicolumn{1}{c}{28.83}                         \\
& \multicolumn{1}{l|}{array\_max\_advanced.c}    & \multicolumn{1}{c|}{22}    &  \multicolumn{1}{c|}{1}      & \multicolumn{1}{c}{4/5}    & \multicolumn{1}{c|}{53.89}       & \multicolumn{1}{c}{5/5}    & \multicolumn{1}{c|}{26.36}     & \multicolumn{1}{c}{5/5}    & \multicolumn{1}{c}{24.17}                         \\
& \multicolumn{1}{l|}{array\_swap.c}             & \multicolumn{1}{c|}{16}    &  \multicolumn{1}{c|}{0}      & \multicolumn{1}{c}{5/5}    & \multicolumn{1}{c|}{13.90}       & \multicolumn{1}{c}{5/5}    & \multicolumn{1}{c|}{3.43}     & \multicolumn{1}{c}{5/5}    & \multicolumn{1}{c}{6.10}                         \\
& \multicolumn{1}{l|}{increment\_arr.c}          & \multicolumn{1}{c|}{16}    &  \multicolumn{1}{c|}{1}      & \multicolumn{1}{c}{0/5}    & \multicolumn{1}{c|}{92.73}       & \multicolumn{1}{c}{4/5}    & \multicolumn{1}{c|}{20.20}     & \multicolumn{1}{c}{0/5}    & \multicolumn{1}{c}{21.57}                         \\
& \multicolumn{1}{l|}{max\_of\_2.c}              & \multicolumn{1}{c|}{13}    &  \multicolumn{1}{c|}{0}      & \multicolumn{1}{c}{5/5}    & \multicolumn{1}{c|}{3.18}       & \multicolumn{1}{c}{5/5}    & \multicolumn{1}{c|}{2.86}     & \multicolumn{1}{c}{5/5}    & \multicolumn{1}{c}{5.27}                         \\ \midrule

\multicolumn{4}{c|}{\textbf{Overall}}          & \multicolumn{1}{c}{27}       & \multicolumn{1}{c|}{100.21}    & \multicolumn{1}{c}{36}    & \multicolumn{1}{c|}{42.04}     & \multicolumn{1}{c}{\cellcolor[HTML]{d5d5d5}\textbf{37}}    & \multicolumn{1}{c}{\cellcolor[HTML]{d5d5d5}\textbf{27.85}}                         \\ \bottomrule
\end{tabular}
\begin{tablenotes}
\scriptsize
\item * \textbf{LoC}: Lines of Code.
\item * \textbf{NoL}: Numbers of Loop.
\end{tablenotes}
\end{threeparttable}
\end{table}
The effectiveness of different approaches on the \textit{frama-c-problems} dataset is reported in Table~\ref{table: Effectiveness of Different Approaches on the frama-c-problems}. Overall, SLD-Spec achieves the best verification performance on this dataset, successfully verifying 37 programs, whereas SpecGen and AutoSpec verify 36 and 27 programs, respectively. In terms of runtime efficiency, SLD-Spec also demonstrates superior performance, with an average running time of 27.85~s, which is significantly lower than AutoSpec’s 100.21~s and also better than SpecGen’s 42.04~s.

Further analysis reveals that for programs without loop constructs (NoL = 0), AutoSpec performs comparably to SLD-Spec and SpecGen in terms of verification success rate. However, when programs contain loops (NoL > 0), AutoSpec’s verification success rate drops significantly. The primary reason is that AutoSpec performs verification immediately after generating loop specifications and directly discards specifications that fail verification when necessary preconditions are missing. This strategy may lead to two adverse consequences: (1) critical specifications may be removed prematurely, preventing them from providing essential support for assertion verification; and (2) when generating function contracts, there are no loop specifications available as references, which further undermines the completeness of the overall specification. In addition, AutoSpec’s iterative generation and verification mechanism increase the number of verification attempts to some extent, thereby significantly increasing the overall runtime. In contrast, SLD-Spec employs LLMs to assess the logical correctness of specifications after generation. With this design, certain loop specifications can be retained as long as they are logically sound, even if the necessary preconditions are temporarily missing. Moreover, SLD-Spec does not rely on iterative mechanisms to repeatedly regenerate specifications, and therefore avoids the runtime overhead introduced by such iterations.

In terms of verification success rate, SpecGen and SLD-Spec exhibit comparable performance. This is because SpecGen treats verification failures as signals for improvement and employs LLMs to repair specifications rather than directly discarding incorrect ones, thereby facilitating the gradual improvement of specification quality. However, due to its one-shot generation approach, SpecGen makes it difficult to precisely control the granularity of specification generation, which may result in the omission of critical specifications. In addition, owing to the inherent randomness in LLM reasoning, SpecGen incurs higher time overhead during the repair phase, primarily arising from the cumulative cost of multiple rounds of specification repair, verification feedback, and model inference.

\subsubsection{RQ2 Effectiveness of complex-func.}
\begin{table*}[]
\centering
\caption{Effectiveness of Different Approaches on the complex-func. }
\label{table: Effectiveness of Different Approaches on the complex-func}
\adjustbox{max width=\textwidth}{
\begin{tabular}{@{}|lll|lll|lll|lll|@{}}
\toprule
\multicolumn{3}{c|}{\textbf{Dataset Information}}  & \multicolumn{3}{c|}{\textbf{AutoSpec}} & \multicolumn{3}{c|}{\textbf{SpenGen}} & \multicolumn{3}{c}{\textbf{SLD-Spec}}  \\ \midrule

\multicolumn{1}{c|}{\multirow{1}{*}{\textbf{Program}}}        &  \multicolumn{1}{c|}{\multirow{1}{*}{\textbf{LoC}}} & \multicolumn{1}{c|}{\multirow{1}{*}{\textbf{NoL}}} & \multicolumn{1}{c}{\multirow{1}{*}{\textbf{PCRSAV}}} & \multicolumn{1}{c}{\multirow{1}{*}{\textbf{NAV}}} & \multicolumn{1}{c|}{\textbf{NPP}} &   \multicolumn{1}{c}{\multirow{1}{*}{\textbf{PCRSAV}}} & \multicolumn{1}{c}{\multirow{1}{*}{\textbf{NAV}}} & \multicolumn{1}{c|}{\textbf{NPP}}  & \multicolumn{1}{c}{\multirow{1}{*}{\textbf{PCRSAV}}} & \multicolumn{1}{c}{\multirow{1}{*}{\textbf{NAV}}} & \multicolumn{1}{c}{\textbf{NPP}} \\ \midrule

\multicolumn{1}{c|}{2-single-loop.c}                     & \multicolumn{1}{c|}{20}    &  \multicolumn{1}{c|}{2}  & \multicolumn{1}{c}{60.74\%}   & \multicolumn{1}{c}{0}    & \multicolumn{1}{c|}{0/5}    & \multicolumn{1}{c}{63.53\%} & \multicolumn{1}{c}{0.4} & \multicolumn{1}{c|}{0/5}    &  \multicolumn{1}{c}{100\%}    & \multicolumn{1}{c}{2}   &  \multicolumn{1}{c}{5/5}                           \\

\multicolumn{1}{c|}{3-single-loop.c}                     & \multicolumn{1}{c|}{27}    &  \multicolumn{1}{c|}{3}  &  \multicolumn{1}{c}{42.21\%}   & \multicolumn{1}{c}{0}    & \multicolumn{1}{c|}{0/5}     & \multicolumn{1}{c}{49.22\%}   & \multicolumn{1}{c}{0}   & \multicolumn{1}{c|}{0/5}      &  \multicolumn{1}{c}{100\%}    & \multicolumn{1}{c}{3}   &   \multicolumn{1}{c}{5/5}                            \\

\multicolumn{1}{c|}{4-single-loop.c}                     & \multicolumn{1}{c|}{34}    &  \multicolumn{1}{c|}{4}  &  \multicolumn{1}{c}{32.84\%}   & \multicolumn{1}{c}{0}    & \multicolumn{1}{c|}{0/5}       & \multicolumn{1}{c}{53.66\%}   & \multicolumn{1}{c}{0.8}   & \multicolumn{1}{c|}{0/5}      &  \multicolumn{1}{c}{100\%}    & \multicolumn{1}{c}{4}   &    \multicolumn{1}{c}{5/5}                            \\ \cmidrule(lr){1-3} \cmidrule(lr){4-12}

\multicolumn{3}{c|}{\textbf{Class Total}}                &  \multicolumn{1}{c}{45.26\%}    & \multicolumn{1}{c}{0}   &   \multicolumn{1}{c|}{0\%}      & \multicolumn{1}{c}{55.47\%}   &  \multicolumn{1}{c}{1.2}   &  \multicolumn{1}{c|}{0\%}       &    \multicolumn{1}{c}{\cellcolor[HTML]{d5d5d5}\textbf{100\%}}    &    \multicolumn{1}{c}{\cellcolor[HTML]{d5d5d5}\textbf{9}}      &    \multicolumn{1}{c}{\cellcolor[HTML]{d5d5d5}\textbf{100\%}}      \\  \midrule

\multicolumn{1}{c|}{only-single-loop-2.c}           & \multicolumn{1}{c|}{26}    &  \multicolumn{1}{c|}{1}  &  \multicolumn{1}{c}{93.67\%}   & \multicolumn{1}{c}{0}    & \multicolumn{1}{c|}{0/5}       & \multicolumn{1}{c}{62.65\%}   & \multicolumn{1}{c}{0}   & \multicolumn{1}{c|}{0/5}      &  \multicolumn{1}{c}{100\%}    & \multicolumn{1}{c}{8}   &   \multicolumn{1}{c}{5/5}                           \\

\multicolumn{1}{c|}{only-single-loop-3.c}           & \multicolumn{1}{c|}{31}    &  \multicolumn{1}{c|}{1}   & \multicolumn{1}{c}{95.45\%}   & \multicolumn{1}{c}{0}    & \multicolumn{1}{c|}{0/5}   & \multicolumn{1}{c}{60.17\%} & \multicolumn{1}{c}{0} & \multicolumn{1}{c|}{0/5}     &  \multicolumn{1}{c}{100\%}    & \multicolumn{1}{c}{9}   &   \multicolumn{1}{c}{5/5}                            \\

\multicolumn{1}{c|}{only-single-loop-4.c}           & \multicolumn{1}{c|}{37}    &  \multicolumn{1}{c|}{1}     & \multicolumn{1}{c}{85.12\%}   & \multicolumn{1}{c}{0}    & \multicolumn{1}{c|}{0/5}   & \multicolumn{1}{c}{62.09\%} & \multicolumn{1}{c}{0} & \multicolumn{1}{c|}{0/5}     &  \multicolumn{1}{c}{99.57\%}    & \multicolumn{1}{c}{8.8}   &   \multicolumn{1}{c}{0/5}                            \\ \cmidrule(lr){1-3} \cmidrule(lr){4-12}

\multicolumn{3}{c|}{\textbf{Class Total}}                &  \multicolumn{1}{c}{91.41\%}    & \multicolumn{1}{c}{0}   &   \multicolumn{1}{c|}{0\%}      & \multicolumn{1}{c}{61.64\%}   &  \multicolumn{1}{c}{0}   &  \multicolumn{1}{c|}{0\%}       &  \multicolumn{1}{c}{\textbf{\cellcolor[HTML]{d5d5d5}99.86\%}}    & \multicolumn{1}{c}{\cellcolor[HTML]{d5d5d5}\textbf{25.8}}   &    \multicolumn{1}{c}{\cellcolor[HTML]{d5d5d5}\textbf{66.67\%}}      \\  \midrule

\multicolumn{1}{c|}{2-loop-1-if-m.c}                       & \multicolumn{1}{c|}{35}     &  \multicolumn{1}{c|}{2}    & \multicolumn{1}{c}{38.25\%}   & \multicolumn{1}{c}{0}    & \multicolumn{1}{c|}{0/5}   & \multicolumn{1}{c}{72.73\%} & \multicolumn{1}{c}{0} & \multicolumn{1}{c|}{0/5}        &  \multicolumn{1}{c}{100\%}    & \multicolumn{1}{c}{3}   &   \multicolumn{1}{c}{5/5}                            \\ 

\multicolumn{1}{c|}{2-loop-1-if-t.c}                       & \multicolumn{1}{c|}{35}     &  \multicolumn{1}{c|}{2}    & \multicolumn{1}{c}{43.85\%}   & \multicolumn{1}{c}{0}    & \multicolumn{1}{c|}{0/5}   & \multicolumn{1}{c}{65.35\%} & \multicolumn{1}{c}{0} & \multicolumn{1}{c|}{0/5}        &  \multicolumn{1}{c}{100\%}    & \multicolumn{1}{c}{2.8}   &   \multicolumn{1}{c}{4/5}                            \\ 

\multicolumn{1}{c|}{2-loop-1-if-b.c}                       & \multicolumn{1}{c|}{35}     &  \multicolumn{1}{c|}{2}    & \multicolumn{1}{c}{51.58\%}   & \multicolumn{1}{c}{1}    & \multicolumn{1}{c|}{0/5}   & \multicolumn{1}{c}{71.72\%} & \multicolumn{1}{c}{0} & \multicolumn{1}{c|}{0/5}        &  \multicolumn{1}{c}{100\%}    & \multicolumn{1}{c}{2.8}   &   \multicolumn{1}{c}{4/5}                            \\ \cmidrule(lr){1-3} \cmidrule(lr){4-12}

\multicolumn{3}{c|}{\textbf{Class Total}}                &  \multicolumn{1}{c}{44.56\%}    & \multicolumn{1}{c}{1}   &   \multicolumn{1}{c|}{0\%}      & \multicolumn{1}{c}{69.93\%}   &  \multicolumn{1}{c}{0}   &  \multicolumn{1}{c|}{0\%}      &    \multicolumn{1}{c}{\textbf{\cellcolor[HTML]{d5d5d5}100\%}}    &    \multicolumn{1}{c}{\cellcolor[HTML]{d5d5d5}\textbf{8.6}}      &    \multicolumn{1}{c}{\cellcolor[HTML]{d5d5d5}\textbf{86.67\%}}      \\  \midrule

\multicolumn{1}{c|}{nested-1.c}              & \multicolumn{1}{c|}{18}    &  \multicolumn{1}{c|}{3}    & \multicolumn{1}{c}{66.91\%}   & \multicolumn{1}{c}{0}    & \multicolumn{1}{c|}{0/5}   & \multicolumn{1}{c}{97.12\%} & \multicolumn{1}{c}{2} & \multicolumn{1}{c|}{5/5}      &  \multicolumn{1}{c}{100\%}    & \multicolumn{1}{c}{1.8}   &   \multicolumn{1}{c}{4/5}                            \\ 

\multicolumn{1}{c|}{nested-2.c}              & \multicolumn{1}{c|}{18}    &  \multicolumn{1}{c|}{3}    & \multicolumn{1}{c}{63.95\%}   & \multicolumn{1}{c}{0}    & \multicolumn{1}{c|}{0/5}   & \multicolumn{1}{c}{100\%} & \multicolumn{1}{c}{2} & \multicolumn{1}{c|}{5/5}      &  \multicolumn{1}{c}{100\%}    & \multicolumn{1}{c}{1.4}   &   \multicolumn{1}{c}{3/5}                            \\ \cmidrule(lr){1-3} \cmidrule(lr){4-12}

\multicolumn{3}{c|}{\textbf{Class Total}}                &  \multicolumn{1}{c}{65.43\%}    & \multicolumn{1}{c}{0}   &   \multicolumn{1}{c|}{0\%}      & \multicolumn{1}{c}{98.56\%}   &  \multicolumn{1}{c}{\cellcolor[HTML]{d5d5d5}\textbf{4}}   &  \multicolumn{1}{c|}{\cellcolor[HTML]{d5d5d5}\textbf{100\%}}       &    \multicolumn{1}{c}{\cellcolor[HTML]{d5d5d5}\textbf{100\%}}    &    \multicolumn{1}{c}{3.2}      &    \multicolumn{1}{c}{70\%}      \\  \midrule

\multicolumn{3}{c|}{\textbf{Number of Passes}}          & \multicolumn{3}{c|}{0}    & \multicolumn{3}{c|}{2}     & \multicolumn{3}{c}{\cellcolor[HTML]{d5d5d5}\textbf{10}}                       \\ \bottomrule
\end{tabular}
}
\end{table*}
The effectiveness of different approaches on the \textit{complex-func} dataset is shown in Table~\ref{table: Effectiveness of Different Approaches on the complex-func}. Overall, SLD-Spec exhibits the most stable and effective performance on this dataset, successfully verifying 10 programs. In contrast, SpecGen succeeds in verifying only 2 programs in the \textit{Nested Loop} category, while AutoSpec fails to complete verification in all categories.

A closer examination of Table~\ref{table: Effectiveness of Different Approaches on the complex-func} shows that the PCRSAV metric differs significantly across approaches, as it is jointly influenced by specification relevance and correctness. AutoSpec adopts a direct deletion strategy for specifications that fail verification, thereby promoting a progressive improvement in the overall correctness of the candidate specifications. However, in programs containing multiple loop structures, AutoSpec’s hierarchical generation mechanism is prone to interference among different loops, leading to the generation of a large number of specifications that are irrelevant to the target loop. As a result, AutoSpec exhibits relatively low PCRSAV values in the \textit{Parallel Single-Path Loop} (45.26\%), \textit{Conditional Enhanced Single-path Loop} (44.56\%), and \textit{Nested Loop} (65.43\%) categories. In contrast, for programs containing only a single loop structure (\textit{Single Multi-Path Loop}), the specifications generated by AutoSpec are more relevant, and its PCRSAV correspondingly improves to 91.41\%.

SpecGen adopts a one-shot specification generation strategy, allowing LLMs to implicitly perform logical separation, and therefore achieves higher overall relevance than AutoSpec. However, because SpecGen relies on LLMs to repair incorrect specifications, it is difficult to guarantee that the retained specifications always satisfy correctness requirements. Moreover, when verification fails due to missing preconditions, SpecGen’s specification mutation mechanism may modify specifications that are originally logically correct, thereby introducing new logical errors and negatively affecting the final PCRSAV.

In contrast, SLD-Spec achieves PCRSAV values close to or reaching 100\% across all categories. This result is mainly attributable to the synergy between its program slicing and logical deletion. Program slicing decomposes complex functions into relatively independent code segments, reducing interference among multiple loops and thereby improving the relevance of the generated specifications. The logical deletion mechanism, in turn, avoids prematurely discarding specifications that temporarily fail verification due to missing preconditions, thus enhancing the overall correctness of the final specification set.

\subsubsection{RQ3 Ablation Study.}
\begin{table*}[]
\renewcommand{\arraystretch}{1.1}
\centering
\caption{Experiment Result of Ablation Study}
\label{table: Experiment Result of Ablation Study}
\adjustbox{max width=\textwidth}{
\begin{tabular}{@{}l|lll|lll|lll|lll@{}}
\toprule
\multicolumn{1}{c|}{\textbf{Dataset}}  & \multicolumn{3}{c|}{\textbf{SLD-Spec w/o both}}   & \multicolumn{3}{c|}{\textbf{SLD-Spec w/o LD}}   & \multicolumn{3}{c|}{\textbf{SLD-Spec w/o PS}}    & \multicolumn{3}{c}{\textbf{SLD-Spec}} \\ \midrule

\multicolumn{1}{c|}{\multirow{1}{*}{\textbf{Program}}}        &  \multicolumn{1}{c}{\multirow{1}{*}{\textbf{PCRSAV}}} & \multicolumn{1}{c}{\multirow{1}{*}{\textbf{NAV}}} & \multicolumn{1}{c|}{\textbf{NPP}} &  \multicolumn{1}{c}{\multirow{1}{*}{\textbf{PCRSAV}}} & \multicolumn{1}{c}{\multirow{1}{*}{\textbf{NAV}}} & \multicolumn{1}{c|}{\textbf{NPP}} &  \multicolumn{1}{c}{\multirow{1}{*}{\textbf{PCRSAV}}} & \multicolumn{1}{c}{\multirow{1}{*}{\textbf{NAV}}} & \multicolumn{1}{c|}{\textbf{NPP}} &  \multicolumn{1}{c}{\multirow{1}{*}{\textbf{PCRSAV}}} & \multicolumn{1}{c}{\multirow{1}{*}{\textbf{NAV}}} & \multicolumn{1}{c}{\textbf{NPP}} \\ \midrule

\multicolumn{1}{c|}{2-single-loop.c}                        &  \multicolumn{1}{c}{69.77\%}    & \multicolumn{1}{c}{1}   &   \multicolumn{1}{c|}{0/5}  &  \multicolumn{1}{c}{85.07\%}    & \multicolumn{1}{c}{1}   &   \multicolumn{1}{c|}{0/5}  &  \multicolumn{1}{c}{98.18\%}    & \multicolumn{1}{c}{2}   &   \multicolumn{1}{c|}{5/5}  &  \multicolumn{1}{c}{100\%}    & \multicolumn{1}{c}{2}   &  \multicolumn{1}{c}{5/5}  \\  

\multicolumn{1}{c|}{3-single-loop.c}                        &  \multicolumn{1}{c}{63.57\%}    & \multicolumn{1}{c}{1}   &   \multicolumn{1}{c|}{0/5}  &  \multicolumn{1}{c}{84.85\%}    & \multicolumn{1}{c}{1}   &   \multicolumn{1}{c|}{0/5}  &  \multicolumn{1}{c}{100\%}    & \multicolumn{1}{c}{1.2}   &   \multicolumn{1}{c|}{0/5}  &  \multicolumn{1}{c}{100\%}    & \multicolumn{1}{c}{3}   &   \multicolumn{1}{c}{5/5}   \\ 

\multicolumn{1}{c|}{4-single-loop.c}                        &  \multicolumn{1}{c}{59.65\%}    & \multicolumn{1}{c}{1}   &   \multicolumn{1}{c|}{0/5}  &  \multicolumn{1}{c}{83.74\%}    & \multicolumn{1}{c}{1}   &   \multicolumn{1}{c|}{0/5}  &  \multicolumn{1}{c}{93.58\%}    & \multicolumn{1}{c}{2.8}   &  \multicolumn{1}{c|}{2/5}   &  \multicolumn{1}{c}{100\%}    & \multicolumn{1}{c}{4}   &    \multicolumn{1}{c}{5/5}   \\ \cmidrule(lr){1-1} \cmidrule(lr){2-13}

\multicolumn{1}{c|}{\textbf{Class Total}}                     &  \multicolumn{1}{c}{64.33\%}    & \multicolumn{1}{c}{3}   &   \multicolumn{1}{c|}{0\%}  &  \multicolumn{1}{c}{84.55\%}    & \multicolumn{1}{c}{3}   &   \multicolumn{1}{c|}{0\%}  &  \multicolumn{1}{c}{97.25\%}    & \multicolumn{1}{c}{6}   &  \multicolumn{1}{c|}{46.67\%}   &  \multicolumn{1}{c}{\cellcolor[HTML]{d5d5d5}\textbf{100\%}}    & \multicolumn{1}{c}{\cellcolor[HTML]{d5d5d5}\textbf{9}}   &    \multicolumn{1}{c}{\cellcolor[HTML]{d5d5d5}\textbf{100\%}}   \\  \midrule

\multicolumn{1}{c|}{only-single-loop-2.c}                   &  \multicolumn{1}{c}{100\%}    & \multicolumn{1}{c}{0}   &   \multicolumn{1}{c|}{0/5}  &  \multicolumn{1}{c}{95.45\%}    & \multicolumn{1}{c}{1}   &   \multicolumn{1}{c|}{0/5}  &  \multicolumn{1}{c}{97.67\%}    & \multicolumn{1}{c}{5.2}   &   \multicolumn{1}{c|}{2/5}  &  \multicolumn{1}{c}{100\%}    & \multicolumn{1}{c}{8}   &   \multicolumn{1}{c}{5/5}   \\ 

\multicolumn{1}{c|}{only-single-loop-3.c}                   &  \multicolumn{1}{c}{100\%}    & \multicolumn{1}{c}{0}   &   \multicolumn{1}{c|}{0/5}  &  \multicolumn{1}{c}{100\%}    & \multicolumn{1}{c}{1}   &  \multicolumn{1}{c|}{0/5}   &  \multicolumn{1}{c}{100\%}    & \multicolumn{1}{c}{2.8}   &   \multicolumn{1}{c|}{1/5}  &  \multicolumn{1}{c}{100\%}    & \multicolumn{1}{c}{9}   &   \multicolumn{1}{c}{5/5}   \\ 

\multicolumn{1}{c|}{only-single-loop-4.c}                   &  \multicolumn{1}{c}{100\%}    & \multicolumn{1}{c}{0}   &   \multicolumn{1}{c|}{0/5}  &  \multicolumn{1}{c}{100\%}    & \multicolumn{1}{c}{1}   &  \multicolumn{1}{c|}{0/5}   &  \multicolumn{1}{c}{98.92\%}    & \multicolumn{1}{c}{2.6}   &  \multicolumn{1}{c|}{0/5}   &  \multicolumn{1}{c}{99.57\%}    & \multicolumn{1}{c}{8.8}   &   \multicolumn{1}{c}{0/5}   \\ \cmidrule(lr){1-1} \cmidrule(lr){2-13} 

\multicolumn{1}{c|}{\textbf{Class Total}}                      &  \multicolumn{1}{c}{\textbf{\cellcolor[HTML]{d5d5d5}100\%}}    & \multicolumn{1}{c}{0}   &   \multicolumn{1}{c|}{0\%}  &  \multicolumn{1}{c}{98.48\%}    & \multicolumn{1}{c}{3}   &   \multicolumn{1}{c|}{0\%}  &  \multicolumn{1}{c}{98.86\%}    & \multicolumn{1}{c}{10.6}   &  \multicolumn{1}{c|}{20\%}   &  \multicolumn{1}{c}{99.86\%}    & \multicolumn{1}{c}{\cellcolor[HTML]{d5d5d5}\textbf{25.8}}   &    \multicolumn{1}{c}{\cellcolor[HTML]{d5d5d5}\textbf{66.67\%}}   \\  \midrule

\multicolumn{1}{c|}{2-loop-1-if-m.c}                      &  \multicolumn{1}{c}{68.42\%}    & \multicolumn{1}{c}{2}   &   \multicolumn{1}{c|}{0/5}  &  \multicolumn{1}{c}{88.51\%}    & \multicolumn{1}{c}{0.8}   &  \multicolumn{1}{c|}{0/5}   &  \multicolumn{1}{c}{99.28\%}    & \multicolumn{1}{c}{2.4}   &  \multicolumn{1}{c|}{3/5}  &  \multicolumn{1}{c}{100\%}    & \multicolumn{1}{c}{3}   &   \multicolumn{1}{c}{5/5}   \\ 

\multicolumn{1}{c|}{2-loop-1-if-t.c}                      &  \multicolumn{1}{c}{77.68\%}    & \multicolumn{1}{c}{0.8}   &   \multicolumn{1}{c|}{0/5}  &  \multicolumn{1}{c}{89.25\%}    & \multicolumn{1}{c}{0.4}   &  \multicolumn{1}{c|}{0/5}   &  \multicolumn{1}{c}{100\%}    & \multicolumn{1}{c}{2.2}   &  \multicolumn{1}{c|}{1/5}  &  \multicolumn{1}{c}{100\%}    & \multicolumn{1}{c}{2.8}   &   \multicolumn{1}{c}{4/5}   \\ 

\multicolumn{1}{c|}{2-loop-1-if-b.c}                      &  \multicolumn{1}{c}{75.25\%}    & \multicolumn{1}{c}{2}   &   \multicolumn{1}{c|}{0/5}  &  \multicolumn{1}{c}{89.90\%}    & \multicolumn{1}{c}{1.4}   &  \multicolumn{1}{c|}{0/5}   &  \multicolumn{1}{c}{100\%}    & \multicolumn{1}{c}{2.2}   &  \multicolumn{1}{c|}{2/5}  &  \multicolumn{1}{c}{100\%}    & \multicolumn{1}{c}{2.8}   &   \multicolumn{1}{c}{4/5}   \\ \cmidrule(lr){1-1} \cmidrule(lr){2-13}

\multicolumn{1}{c|}{\textbf{Class Total}}                      &  \multicolumn{1}{c}{73.78\%}    & \multicolumn{1}{c}{4.8}   &   \multicolumn{1}{c|}{0\%}  &  \multicolumn{1}{c}{89.22\%}    & \multicolumn{1}{c}{2.6}   &   \multicolumn{1}{c|}{0\%}  &  \multicolumn{1}{c}{99.76\%}    & \multicolumn{1}{c}{6.8}   &  \multicolumn{1}{c|}{40\%}   &  \multicolumn{1}{c}{\cellcolor[HTML]{d5d5d5}\textbf{100\%}}    & \multicolumn{1}{c}{\cellcolor[HTML]{d5d5d5}\textbf{8.6}}   &    \multicolumn{1}{c}{\cellcolor[HTML]{d5d5d5}\textbf{86.67\%}}   \\  \midrule

\multicolumn{1}{c|}{nested-1.c}                        &  \multicolumn{1}{c}{79.52\%}    & \multicolumn{1}{c}{0}   &   \multicolumn{1}{c|}{0/5}  &  \multicolumn{1}{c}{100\%}    & \multicolumn{1}{c}{0}   &  \multicolumn{1}{c|}{0/5}  &  \multicolumn{1}{c}{100\%}    & \multicolumn{1}{c}{0.6}   &   \multicolumn{1}{c|}{1/5}  &  \multicolumn{1}{c}{100\%}    & \multicolumn{1}{c}{1.8}   &   \multicolumn{1}{c}{4/5}   \\ 

\multicolumn{1}{c|}{nested-2.c}                        &  \multicolumn{1}{c}{74.63\%}    & \multicolumn{1}{c}{0}   &   \multicolumn{1}{c|}{0/5}  &  \multicolumn{1}{c}{100\%}    & \multicolumn{1}{c}{0}   &  \multicolumn{1}{c|}{0/5}  &  \multicolumn{1}{c}{100\%}    & \multicolumn{1}{c}{0.4}   &   \multicolumn{1}{c|}{0/5}  &  \multicolumn{1}{c}{100\%}    & \multicolumn{1}{c}{1.4}   &   \multicolumn{1}{c}{3/5}   \\ \cmidrule(lr){1-1} \cmidrule(lr){2-13}

\multicolumn{1}{c|}{\textbf{Class Total}}                      &  \multicolumn{1}{c}{77.08\%}    & \multicolumn{1}{c}{0}   &   \multicolumn{1}{c|}{0\%}  &  \multicolumn{1}{c}{\textbf{\cellcolor[HTML]{d5d5d5}100\%}}    & \multicolumn{1}{c}{0}   &   \multicolumn{1}{c|}{0\%}  &  \multicolumn{1}{c}{\textbf{\cellcolor[HTML]{d5d5d5}100\%}}    & \multicolumn{1}{c}{1}   &  \multicolumn{1}{c|}{10\%}   &    \multicolumn{1}{c}{\cellcolor[HTML]{d5d5d5}\textbf{100\%}}    &    \multicolumn{1}{c}{\cellcolor[HTML]{d5d5d5}\textbf{3.2}}      &    \multicolumn{1}{c}{\cellcolor[HTML]{d5d5d5}\textbf{70\%}}   \\  \midrule

\multicolumn{1}{c|}{\textbf{Number of Passes}}                   &  \multicolumn{3}{c|}{0}     &  \multicolumn{3}{c|}{0}     &  \multicolumn{3}{c|}{8}    &    \multicolumn{3}{c}{\cellcolor[HTML]{d5d5d5}\textbf{10}}  \\ \bottomrule
\end{tabular}
}
\end{table*}
To evaluate the impact of program slicing and logical deletion on the performance of SLD-Spec, we compare it with three variants, as shown in Table~\ref{table: Experiment Result of Ablation Study}. Compared with the baseline variant that includes neither mechanism (SLD-Spec w/o both), the variant that incorporates only program slicing (SLD-Spec w/o LD) effectively improves specification relevance by decomposing programs into smaller slices and generating specifications for each slice, thereby significantly increasing PCRSAV. On the other hand, the variant that incorporates only logical deletion (SLD-Spec w/o PS) not only improves relevance by using LLMs to eliminate some irrelevant specifications, but also retains specifications that are logically correct yet fail verification due to missing preconditions during the verification phase. These specifications are subsequently refined by correcting their contextual dependencies through further reasoning, which enhances specification correctness. As a result, this variant achieves comprehensive improvements in PCRSAV, NAV, and NPP, and ultimately succeeds in fully verifying eight programs. However, because it relies on LLMs to identify and remove irrelevant specifications, its stability is still affected to some extent. The full SLD-Spec, which combines both program slicing and logical deletion, achieves PCRSAV values close to or reaching 100\% across all categories and successfully verifies ten programs. Program slicing reduces interference between slices and enables the generation of more complete specifications, while logical deletion preserves logically correct specifications. Consequently, SLD-Spec attains the highest performance on both NAV and NPP, significantly outperforming all three variants.

\section{Threats to Validity}
Although SLD-Spec demonstrates strong empirical performance, its effectiveness is subject to the following three main threats to validity. First, \textbf{the risk of data leakage}. Since the frama-c-problems dataset has been available for a relatively long period, its contents may have been included in the training data of some LLMs, thereby introducing potential data leakage. In such cases, the absolute performance metrics of all compared methods may be overestimated. However, because all methods are evaluated under the same experimental settings, data leakage is unlikely to introduce systematic bias in the relative comparisons among different methods. In contrast, the complex-func dataset is newly constructed in this study and has not been publicly released prior to this work, which partially mitigates the risk of data leakage. Second, \textbf{the reliability risk of logical deletion}. This step relies on LLMs to assess the logical correctness of candidate specifications. Due to the inherent non-determinism of LLMs, logically correct specifications may be incorrectly judged as incorrect and removed, and vice versa. To mitigate this risk, we introduce a multi-sampling voting mechanism, in which LLMs are invoked multiple times on the same batch of specifications, and the final decision is made based on majority voting. This approach helps reduce the bias of individual judgments and improves the stability of the logical deletion process. Third, \textbf{the limited dataset size}. In our experiments, the frama-c-problems dataset contains 51 programs, and the complex-func dataset contains 11 programs. Although the effectiveness of SLD-Spec has been validated on these datasets, their scale may still be insufficient to fully capture the complexity and diversity of real-world software systems, which may limit the generalizability of the proposed method. Nevertheless, the two datasets cover a variety of program categories (eight categories) and different control-flow patterns (four types), providing a meaningful basis for evaluating our approach to some extent. To further mitigate this threat, we plan to construct larger and more diverse datasets in future work to better reflect practical software development scenarios.

\section{Related Work}
\subsection{Traditional Specification Generation Approaches}
Early related work can generally be categorized into three classes: static analysis (SA)-based~\cite{Gosain-01}, dynamic detection (DD)-based~\cite{Ball-01}, and machine learning (ML)-based~\cite{Pandey-01} approaches. SA-based approaches~\cite{Albarghouthi-01,Alur-01,Engler-01,Nguyen-01,Ramanathan-01,Shoham-01,Weimer-01} analyze source code through techniques such as static inspection, abstract interpretation, or inductive reasoning to identify execution paths and relationships among variables, and then generate specifications based on parameterized templates. For example, Alur et al.~\cite{Alur-01} proposed an automated method based on predicate abstraction and the L* learning algorithm to extract safe behavioral interface specifications from Java class code. However, the scalability of such approaches is often limited when confronted with complex data structures. DD-based approaches~\cite{Ammons-01,Ernst-01,Gabel-01,Lee-01,Lemieux-01,Padhi-01,Reger-01,Whaley-01,Yang-02} rely on actual program executions. By executing test cases and tracing runtime outputs and intermediate states, they capture path and data-flow information and generate specifications that reflect observed program behaviors. For instance, Michael et al.~\cite{Ernst-01} developed Daikon, a dynamic invariant detection tool that automatically infers likely invariants by observing variable values during program execution, thereby supporting program comprehension and verification. However, the effectiveness of these approaches largely depends on test case coverage. ML-based approaches~\cite{Krismayer-01,Liu-01,Mrowca-01,Ryan-01,Si-01,Yao-01,Yu-01} leverage neural networks to learn program behavior patterns from large numbers of execution traces, infer variable dependencies and potential execution paths, and predict program behavior under specific input conditions for specification generation. For example, Si et al.~\cite{Si-01} proposed CODE2INV, an end-to-end approach based on graph neural networks and reinforcement learning that automatically generates loop invariants through a multi-step decision-making process. However, such approaches typically require large amounts of high-quality training data to achieve strong generalization.

\subsection{LLM-based Specification Generation Approaches}
In recent years, with the rise of LLMs, a growing body of research has begun to explore their application in software engineering. LLM-based specification generation approaches~\cite{Chakraborty-01,Fan-01,Kamath-01,Ma-01,Pei-01,Wen-01,Wu-01,Xie-01} leverage the strong comprehension and generation capabilities of LLMs to overcome the limitations of traditional specification generation methods. For example, Xie et al.~\cite{Xie-01} were among the first to evaluate the capability of LLMs to generate specifications from software comments or documentation. In contrast, our work focuses on generating specifications directly from program source code rather than software documentation. Pei et al.~\cite{Pei-01} investigated the capability of fine-tuned LLMs to generate program invariants, whereas our work instead employs off-the-shelf LLMs for specification generation without retraining large models. Kamath et al.~\cite{Kamath-01} designed the earliest LLM-based automated specification generation framework, which queries LLMs to generate specifications for an entire program and then checks their correctness using verification tools. Building on this framework, Wen et al.~\cite{Wen-01} proposed AutoSpec, which leverages static analysis to hierarchically decompose programs and incrementally generate specifications in a bottom-up manner. Ma et al.\cite{Ma-01} introduced an LLM-based technique for automatically generating specifications for Java programs, employing a conversational paradigm to guide LLMs in producing appropriate specifications for a given program and using heuristic algorithms to select mutated specifications. Wu et al.~\cite{Wu-01} proposed LEMUR, a proof calculus that combines LLMs with automated reasoners. Compared with existing studies, SLD-Spec focuses on addressing critical deficiencies in current specification generation frameworks, namely semantic interference during the generation phase and overly aggressive elimination and unreliable filtering during the verification phase.

\section{Conclusion}
We propose SLD-Spec, an LLM-based specification generation approach enhanced by program slicing and logical deletion. Our approach decomposes functions into smaller slices and uses LLMs to filter logically correct specifications. Experimental results show that SLD-Spec outperforms other baselines in both verification capability and efficiency, thereby effectively improving the program verification success rate.

%
%
%
\bibliographystyle{splncs04}
\bibliography{references}
%




\appendix
\section{Case Studies of Verification Failure}
\label{Case Studies of Verification Failure}
\begin{figure}
    \centering
    \includegraphics[width=\textwidth]{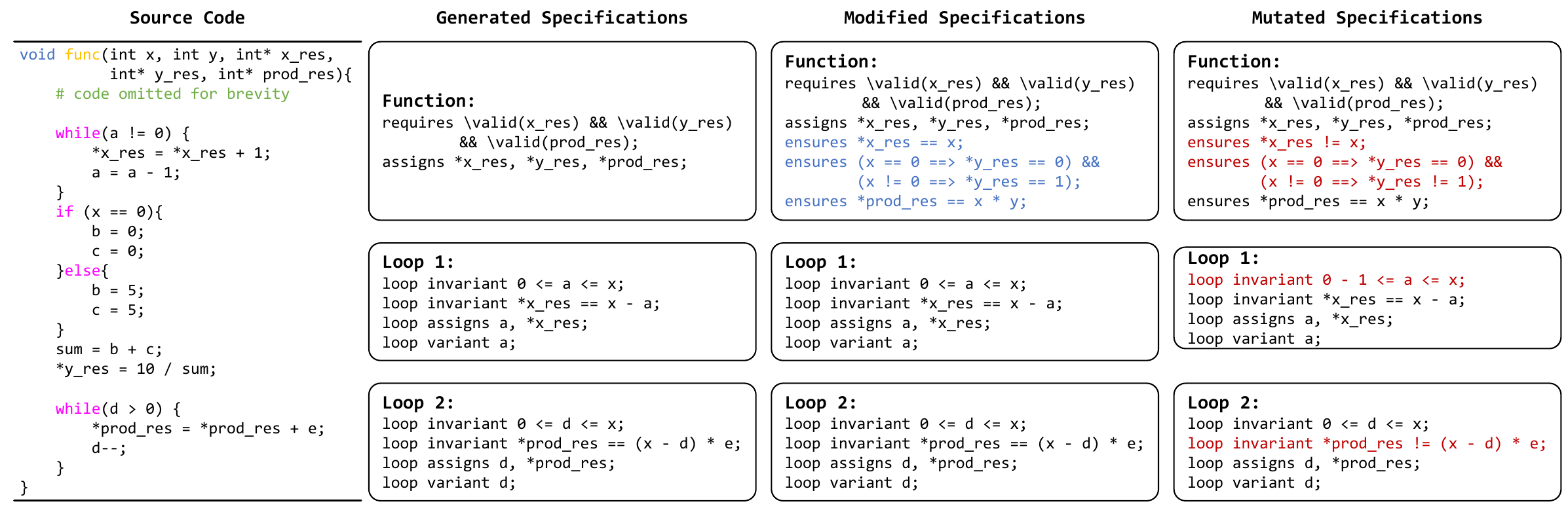}
    \caption{Case 1.}
    \label{pic: case 1}
\end{figure}
\subsection{Case 1: Complex Control Flow Causes Verification Failures in SpecGen}
We demonstrate how complex control flow can lead to verification failures in SpecGen, and how its mutation mechanism may incorrectly modify logically correct specifications. As shown in Fig.~\ref{pic: case 1}, given the source code on the left, SpecGen generates an initial set of specifications. Although SpecGen produces correct and complete specifications for both loops, the structural complexity of the code results in an incomplete function contract. During the subsequent repair phase, SpecGen leverages the capabilities of LLMs to supplement some of the missing postconditions (shown in blue in Fig.~\ref{pic: case 1}). However, it consistently fails to generate two crucial preconditions: {\consolaslike{requires \textbackslash separated(prod\_res, y\_res, x\_res)}} and {\consolaslike{requires x >= 0}}. The absence of these specifications causes other specifications to fail verification, even though they are logically correct. Finally, SpecGen mutates the specifications that fail verification, leading to incorrect modifications of originally correct specifications (shown in red in Fig.~\ref{pic: case 1}), thereby further exacerbating the errors.

\begin{figure}
    \centering
    \includegraphics[width=\textwidth]{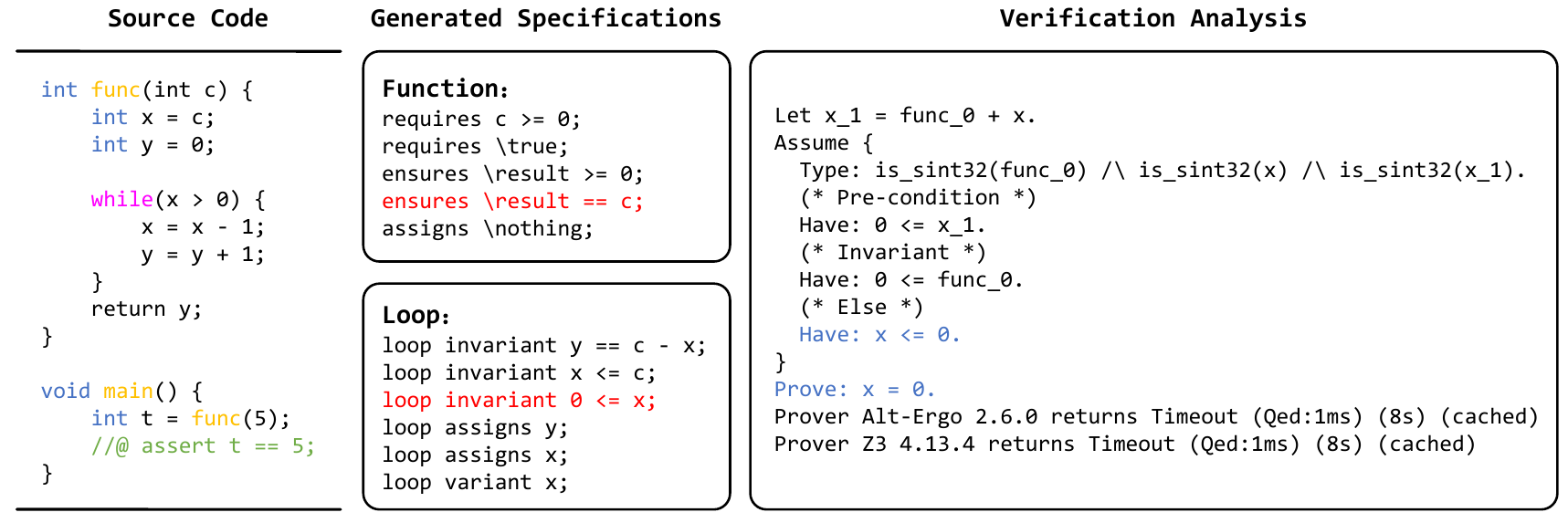}
    \caption{Case 2.}
    \label{pic: case 2}
\end{figure}

\subsection{Case 2: Immediate Verification After Generation Causes Failures in AutoSpec}
The example in Fig.~\ref{pic: case 2} shows that even for simple programs, AutoSpec’s strategy of immediately verifying specifications after generation can lead to verification failures. Specifically, for the source code on the left, AutoSpec first generates specifications for the loop. However, since the precondition {\consolaslike{requires c >= 0}} has not yet been generated at this stage, the specification {\consolaslike{loop invariant 0 <= x}} fails during verification and is therefore removed. Subsequently, after function contracts are generated, the postcondition {\consolaslike{ensures \textbackslash result == c}} also fails to verify, which in turn causes the assertion in the source code to fail verification. The verification analysis of this postcondition, shown on the right, indicates that to prove {\consolaslike{\textbackslash result == c}} (i.e., {\consolaslike{y == c}}), it is necessary to establish that the loop exit condition {\consolaslike{x == 0}} holds, given the existing {\consolaslike{loop invariant y == c - x}}. Based on the available specifications, the verification tool can only derive {\consolaslike{x <= 0}}; when attempting to further prove {\consolaslike{x == 0}}, it times out, leading to verification failure. In principle, this issue could be resolved by constraining the lower bound of x, namely by introducing the specification {\consolaslike{0 <= x}}, which would allow {\consolaslike{x == 0}} to be proven. However, this specification was removed in an earlier phase, ultimately causing the entire program to fail verification.

\begin{figure}
    \centering
    \includegraphics[width=0.4\textwidth]{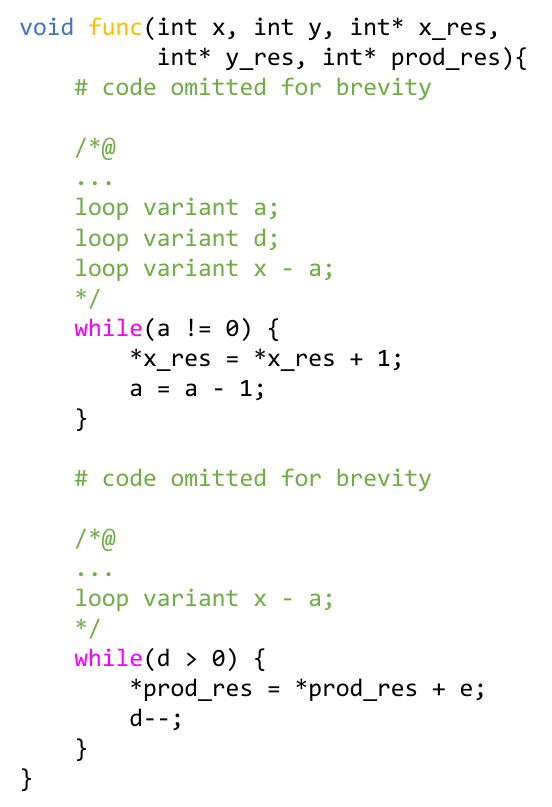}
    \caption{Case 3.}
    \label{pic: case 3}
\end{figure}

\subsection{Case 3: Limitations of Using Verification Tools to Filter Specifications}
Fig.~\ref{pic: case 3} illustrates a case in which using a verification tool to filter candidate specifications leads to verification failure; some irrelevant code and specifications are omitted for clarity. As shown in Fig.~\ref{pic: case 3}, AutoSpec generates three different {\consolaslike{loop variant}} clauses for the first loop and attempts to eliminate incorrect candidates using the verification tool. However, in this situation, the verification tool does not determine correctness through mathematical reasoning; instead, it treats the earliest {\consolaslike{loop variant}} clause as a syntactic error. AutoSpec then removes the corresponding specification based on this analysis. After two rounds of verification, the first loop ultimately retains an incorrect clause, namely {\consolaslike{loop variant x - a}}. This error further propagates to the specification generation of the second loop, causing both loops to ultimately retain incorrect {\consolaslike{loop variant}} clauses.

\subsection{Case 4: The Programs Failing Verification by SLD-Spec on the Complex-Func Dataset}
In the \textit{Single Multi-Path Loop} category of the \textit{complex-func} dataset, there is a program, \textit{only-single-loop-4.c}, that SLD-Spec fails to verify. This program contains a single loop whose body includes four distinct if branches. Through detailed analysis, we found that SLD-Spec is not inherently incapable of generating correct and complete specifications for this program; rather, proving some of the generated specifications requires several hundred seconds, which necessitates configuring an extremely long timeout for the verification tool. However, SLD-Spec takes the union of results from multiple generation runs as the set of candidate specifications, most of which are incorrect or repetitive. If a long timeout is used for the verification tool, a substantial amount of time is wasted attempting to verify these specifications. Consequently, under the current configuration, SLD-Spec is unable to successfully verify this program. To validate this conclusion, we manually selected the 38th retained result from the third independent run of SLD-Spec for this program as the basis and set the verification tool’s timeout to 300~s. After a single verification and manual elimination of incorrect specifications, the program was ultimately verified successfully.
\end{document}